\documentclass[onecolumn,amsmath,amssymb,11pt]{revtex4}

\def\gsim{ \lower .75ex \hbox{$\sim$} \llap{\raise .27ex \hbox{$>$}} }

\def\lsim{ \lower .75ex \hbox{$\sim$} \llap{\raise .27ex \hbox{$<$}} }

\usepackage{graphicx}
\usepackage{dcolumn}
\usepackage{bm}
\input epsf
\def\gsim{ \lower .75ex \hbox{$\sim$} \llap{\raise .27ex \hbox{$>$}} }
\def\lsim{ \lower .75ex \hbox{$\sim$} \llap{\raise .27ex \hbox{$<$}} }

\usepackage{graphicx}
\usepackage{dcolumn}
\usepackage{bm}

\newcommand{\nn}{\nonumber}
\newcommand{\be}{\begin{equation}}
\newcommand{\ee}{\end{equation}}
\newcommand{\bea}{\begin{eqnarray}}
\newcommand{\eea}{\end{eqnarray}}

\def\a{\alpha}

\def\d{\delta}

\def\s{\sigma}
\def\sb{\bar{\sigma}}
\def\e{\epsilon}

\def\k{\kappa}

\def\th{\theta}
\def\tb{\bar{\theta}}

\def\r{\rho}
\def\rb{\bar{\rho}}
\def\s{\sigma}

\def\t{\tau}

\def\x{\xi}
\def\z{\zeta}

\def\pt{\partial}
\def\half{\frac{1}{2}}
\def\besi{\bar{e}_{\s}}
\def\bes{\bar{e}_{s}}

\def\p{\phi}
\def\tackr{&\!\!\!}
\def\tackl{&\!\!\!}
\def\sib{\bar{\sigma}}
\def\thetab{\bar{\theta}}

\def\uc{u}
\def\Dc{D}
\def\L{{\cal L}}

\def\p{\varphi}
\def\I{I}

\def\hc{h}
\def\sigmad{\dot \sigma}
\def\sif{{\delta \sigma}}
\def\sis{{\delta \sigma}^{(2)}}
\def\sf{{\delta s}}
\def\ss{{\delta s}^{(2)}}

\def\rhos{{\delta \rho}^{(2)}}

\def\sp{\sigma^{_\perp}}

\begin{document}

\title{Multifield Cosmological Perturbations at Third Order and the \\ Ekpyrotic Trispectrum}

\author{Jean-Luc Lehners$^\ddag$ and S\'{e}bastien Renaux-Petel$^\S$}
\affiliation{$^\ddag$ Princeton Center for Theoretical Science,
Princeton University, Princeton, NJ 08544 USA \\ $^\S$ APC (UMR
7164, CNRS, Universit\'e Paris 7), 10 rue Alice Domon et
L\'eonie Duquet, 75205 Paris Cedex 13, France }

\begin{abstract}
Using the covariant formalism, we derive the equations of
motion for adiabatic and entropy perturbations at third order
in perturbation theory for cosmological models involving two
scalar fields. We use these equations to calculate the
trispectrum of ekpyrotic and cyclic models in which the density
perturbations are generated via the entropic mechanism. In
these models, the conversion of entropy into curvature
perturbations occurs just before the big bang, either during
the ekpyrotic phase or during the subsequent kinetic energy
dominated phase. In both cases, we find that the non-linearity
parameters $f_{NL}$ and $g_{NL}$ combine to leave a very distinct
observational imprint.
\end{abstract}


\maketitle
\tableofcontents{}

\section{Introduction and summary of results}

The cosmic microwave background radiation (CMB) provides us
with a staggering amount of information about the early
universe. In fact, its exploration has only recently begun in
detail with the WMAP satellite, but with the successful launch
of the Planck satellite, we can hope to get yet more precise
maps of the microwave sky in the near future. At the same time, surveys of the large-scale structure of the universe are becoming increasingly detailed, to the point where they can rival the precision of the CMB maps, albeit with more complicated, and less understood, physics involved. The analysis of the statistics of
density fluctuations is thus becoming a strong
discriminator of competing models of the early universe.

An increasingly important aspect is non-gaussianity
\cite{Komatsu:2009kd}, {\it i.e.} the deviation from perfect
gaussian statistics. Indeed, there are many models of the early
universe that, at the linear level, give degenerate predictions
({\it e.g.} the large class of inflationary models
\cite{Linde:2007fr} and ekpyrosis \cite{Lehners:2008vx}) -
hence the importance of studying higher correlation functions.
As we go to higher $n$-point functions, the degeneracy
typically tends to be broken, as has been demonstrated by the
study of the bispectrum in recent years. Clearly,
the larger the number of distinct observations that a
cosmological model is in agreement with, the more plausible it
becomes, and this constitutes the primary reason for extending
the analysis to the trispectrum, {\it i.e.} the Fourier
transform of the 4-point function.

Even though we are looking at what appear to be increasingly
small details of the CMB maps, the implications can be
far-reaching, since different models can give rise to vastly
different pictures of the universe in terms of its long-time
evolution and its appearance on scales larger than the current
horizon - in this vein one should for example compare eternal
inflation \cite{Guth:2007ng} and the phoenix universe
\cite{Lehners:2008qe}. Hence it is of crucial importance to
understand the predictions of the various models, ideally
before the measurements are made, so that these measurements
can be interpreted within a suitably developed theoretical
framework.

In this paper, we extend to third order in perturbation theory
the derivation of the equations of motion needed to calculate
the non-gaussianity of two-scalar-field cosmological models.
For this, we use the covariant formalism, as it provides a
conceptually unambiguous path to the definition of higher-order
perturbations and the derivation of their equations of motion
(for some of our examples, the background evolution is
complicated, so that we cannot easily use the $\d N$
formalism). The resulting equations are generally applicable,
and we have used them here to calculate the trispectrum of
perturbations generated by ekpyrotic and cyclic cosmological
models. For certain inflationary models, the trispectrum has
already been calculated using the $\delta N$ formalism
\cite{Seery:2006js,Seery:2006vu,Byrnes:2006vq}, and, in future
work, it would certainly be of interest to use the formalism
developed here in order to check and extend these results to
cases where the $\delta N$ formalism is less readily
applicable.

Ekpyrotic models constitute a case of particular interest
regarding non-gaussianities, since the cosmological
perturbations are generated in the presence of a steep
potential. Hence, we are automatically in a regime where we
expect significant self-interactions of the scalar fields, and
hence significantly non-gaussian distributions of temperature
fluctuations. This is in stark contrast to simple inflationary
models \cite{Acquaviva:2002ud,Maldacena:2002vr}, where the
flatness of the potential implies that the inflaton is
practically massless, with negligible self-interactions. In
more complicated inflationary models, such as the curvaton
model \cite{Lyth:2001nq,Lyth:2002my}, significant
non-gaussianities can be developed due to the interactions of
various scalar fields, but with the drawback that the range of
possible values for the non-linearity parameters is difficult
to pin down \footnote{In this paper, we focus on the so-called
``local'' form of the non-linearity parameters, which is the
relevant form to consider in the presence of canonical kinetic
terms. Non-canonical kinetic terms usually imply that different
shape functions in momentum space are more relevant; the
trispectrum for models involving non-canonical kinetic terms
has recently been computed \cite{Chen:2009bc,Arroja:2009pd} -
see these papers as well as \cite{InPreparation2} for
details.}. The beauty of the ekpyrotic models is that, despite
the fact that the potential has not been derived
microphysically, and that we therefore must use a parameterized
form for the potential, the results are nevertheless
surprisingly predictive.

In this paper, we restrict our
attention to those models in which the curvature perturbations
are generated via the entropic mechanism \cite{Lehners:2007ac},
as this is the most robust mechanism known to date. In this
mechanism, nearly scale-invariant entropy perturbations are
generated first, and are subsequently converted into curvature
perturbations via a bending of the trajectory in scalar field
space just before the big bang. Such a bending of the
trajectory is generically expected in various contexts
\cite{Lehners:2007nb,Koyama:2007mg,Buchbinder:2007ad}. Broadly
speaking, there are two limiting cases that are of special
interest: the first is where the bending occurs after the
ekpyrotic phase, during the approach to the big crunch. In this
case, which is the natural case in cyclic models, the kinetic
energy of the scalar fields is the dominant contribution to the
total energy density while the conversion takes place, and
hence we call this case ``kinetic conversion''. The other limiting case is where the conversion occurs during
the ekpyrotic phase, and we call this ``ekpyrotic conversion''.

\smallskip

Postponing all details to sections \ref{sectionNonlinearity} and
\ref{sectionEkpyrotic}, we simply state our main
results here. We define the local non-linearity parameters $f_{NL}$ and $g_{NL}$ via an expansion of the curvature perturbation $\zeta$ in terms of its linear, gaussian part $\zeta_L,$
\be \z = \z_L + \frac{3}{5} f_{NL} \z_L^2 + \frac{9}{25} g_{NL} \z_L^3. \ee During the ekpyrotic phase, we adopt the following parametrization of the
potential: \be V_{ek}=-V_0
e^{\sqrt{2\e}\s}[1 +\e s^2 +\frac{\k_3}{3!}\e^{3/2} s^3
+\frac{\k_4}{4!}\e^2 s^4+\cdots],\ee where we expect $\k_3,\k_4
\sim {\cal O}(1)$ and where $\e \sim {\cal O}(10^{2})$ is
related to the ekpyrotic equation of state $w_{ek}$ via
$\e=3(1+w_{ek})/2.$ $\s$ denotes the adiabatic direction, {\it
i.e.} the direction tangent to the scalar field space
trajectory, while $s$ denotes the ``entropy'' direction, {\it
i.e.} the direction perpendicular to the background trajectory.
During the ekpyrotic phase, we are able to solve for the
entropy perturbation, with the result that \be \d s= \d s_L +
\frac{\k_3 \sqrt{\e}}{8}\d s_L^2 + \e
(\frac{\k_4}{60}+\frac{\k_3^2}{80}-\frac{2}{5})\d s_L^3, \ee
with the linear, gaussian part $\d s_L$ being inversely
proportional to time $t$.

Then, for {\em kinetic conversions} lasting on the
order of one e-fold of contraction of the scale factor, we find
the following approximate fitting formula for the
third order non-linearity parameter $g_{NL}$ (and we include $f_{NL}$ for completeness \cite{Lehners:2007wc,Lehners:2008my})
: \bea f_{NL} &\sim& \frac{3}{2} \, \k_3 \sqrt{\e} +5 \\
g_{NL} &\sim& 100 \,
\e(\frac{\k_4}{60}+\frac{\k_3^2}{80}-\frac{2}{5}). \eea These
simple formulae accurately reflect the scaling with
$\e,\k_3,\k_4$ of the non-linearity parameters. But crucially,
note that when $\k_3$ and $\k_4$ are small, $g_{NL}$ is always
negative. Hence, even though $f_{NL}$ is small in that case,
$g_{NL}$ is negative and typically of order a few thousand, so
that any accidental degeneracy at the level of $f_{NL}$ between
simple inflationary and cyclic models is very likely to be
broken at the level of the trispectrum. More generally, unless
$|f_{NL}|$ turns out to be quite large, one would typically
expect $g_{NL}$ to be negative, as obtaining a positive
$g_{NL}$ in that case would require unnaturally large values of
$\k_4.$

For {\em ekpyrotic conversions}, it is easier to perform the
analysis in terms of the orthogonal fields $\phi_1$ and
$\phi_2,$ with $\phi_2$ being the direction of the trajectory
after the bending ($\phi_1$ and $\phi_2$ are related to $\s$
and $s$ by a rotation in field space). Then, if the potential
is expressed as \be V_{ek} = - V_1 e^{-c_1 \phi_1} -V_2 e^{-c_2
\phi_2},\ee we find the following approximate formulae for the
non-linearity parameters (where $f_{NL}$ was first calculated in \cite{Koyama:2007if}):
\bea f_{NL} &\sim& -\frac{5}{12}c_1^2 \\
 g_{NL} &\sim& \frac{25}{108}c_1^4. \eea
Here $f_{NL}$ is always negative while $g_{NL}$ is always
positive, and both scale rather fast with increasing $c_1$.
Since we must have $c_1\gtrsim 8$ in order not to conflict with
spectral index measurements, we can expect future observations
to be unambiguous in their support or refutation of these
models, and current limits on $f_{NL}$ are already putting some
strain on this particular mode of conversion.

\bigskip

The plan of this paper is as follows: we start with a short
summary of the covariant formalism, and of how it has been used
to derive the equations of cosmological perturbations up to
second order in perturbation theory. Then we extend these
results to third order, and derive the equations of motion
relevant for calculating the trispectrum of cosmological
perturbations. In section \ref{sectionNonlinearity}, we define
the non-linearity parameters that are typically used to quote
observational results, and that we wish to calculate. Then, in
section \ref{sectionEkpyrotic} we apply our equations to
ekpyrotic models, and numerically solve them for the cases of
kinetic and ekpyrotic conversion. We end with a discussion of
our results. The appendix contains a collection of useful
formulae.

\section{Covariant formalism and perturbation theory up to second order}
\label{sectionCovariant}

Motivated by the calculation of the primordial trispectrum in
ekpyrotic models, we study the cosmological fluctuations of a
system of two scalar fields at third order in perturbation
theory. We adopt the 1+3 covariant formalism developed by
Langlois and Vernizzi
\cite{Langlois:2005ii,Langlois:2005qp,Langlois:2006iq,Langlois:2006vv},
along the lines of earlier works by Ellis and Bruni
\cite{Ellis:1989jt} and Hawking \cite{Hawking:1966qi} (see also
\cite{Bruni:1991kb}). Applications of this formalism to a wide
variety of cosmological contexts can be found in
\cite{Tsagas:2007yx}. It is particularly well suited to analyze
fluctuations of the system we consider, as it provides a
geometric and, most importantly, a fully non-perturbative
description of spacetimes that are distorted away from the
standard FLRW form.

The idea is to define perturbations via
spatial projections of various scalar quantities, such as the
energy density or the integrated expansion along a time-like
curve. These spatial projections (covectors) vanish for an
exact homogeneous and isotropic FLRW spacetime, and hence they
naturally play the role of `perturbations', even though they
can be defined non-perturbatively. The formalism allows one to
derive simple, all-orders, equations of motion for these
covectors, which, after choosing a convenient coordinate basis,
can be expanded up to the desired order in perturbation theory.
Also, the formalism leads to a clear and natural decomposition
of fluctuations into adiabatic and entropy components, again in
a fully non-perturbative way.

In this section, we give an overview of the covariant formalism
and its application to the study of a system of two scalar
fields, up to second order in perturbation theory, as done in
\cite{Langlois:2006vv} (see also \cite{RenauxPetel:2008gi} for
an extension to two-field models with non-standard kinetic terms of the form studied in \cite{Langlois:2008mn,Gao:2008dt}). This provides the
necessary groundwork for our investigation of third-order
perturbation theory in the following section. We will also
focus on the large scale limit, {\it i.e} we only consider the
leading-order terms in an expansion in spatial gradients. In
this respect, our formalism is similar to the one developed in
\cite{Rigopoulos:2005xx}.

Our analysis is quite general: we do not specify an underlying
potential and do not use approximations other than the large
scale limit. In particular, although we will apply it to a
particular scenario of a contracting universe, it is equally
well suited to the study of cosmological perturbations
generated during a period of inflation.

\subsection{Covariant formalism}

A key role in the covariant splitting of space and time in the
1+3 covariant formalism is played by the {\it a priori} arbitrary
unit timelike vector $u^a= d x^a/d \tau$ ($u_a u^a =-1$),
defining a congruence of cosmological observers. The spatial
projection tensor orthogonal to the four-velocity $u^a$ is
provided by \be h_{ab}\equiv g_{ab}+u_a u_b, \quad \quad
(h^{a}_{\ b} h^b_{\ c}=h^a_{\ c}, \quad h_a^{\ b}u_b=0). \ee
where $g_{ab}$ is the cosmological metric. To describe the time
evolution, the covariant definition of the time derivative is
then the {\em Lie derivative} with respect to $u^a$, defined
for a generic covector $Y_a$ by (see e.g. \cite{Wald:1984rg})
\begin{equation}
\dot Y_a\equiv\L_u Y_a \equiv u^b \nabla_b Y_a + Y_b \nabla_a u^b,
\label{Lie}
\end{equation}
and denoted by {\em a dot}. For scalar  quantities, one simply
has \be \dot f = u^b \nabla_b f. \ee To describe perturbations
in the covariant approach, we consider the projected covariant
derivative orthogonal to the four-velocity $\uc^a$, denoted by
$\Dc_a$. For a generic tensor, the definition is \be D_a
T_{b\dots}^{\ c\dots}\equiv h_a^{\ d}h_b^{\ e}\dots h^{\
c}_f\dots\nabla_d T_{e\dots}^{\ f\dots}. \ee In particular,
when focussing on a scalar quantity $f$, this reduces to \be
\Dc_a f\equiv h_a^{\ b} \nabla_b f = \partial_a f + u_a \dot f
\, . \label{def_Daf} \ee Geometrical properties of the
congruence are defined via the decomposition \be \nabla_b
u_a=\sigma_{ab}+\omega_{ab}+{1\over 3}\Theta h_{ab}-a_a u_b,
\label{decomposition} \ee with the (symmetric) shear tensor
$\sigma_{ab}$ and the (antisymmetric) vorticity  tensor
$\omega_{ab}$; the volume expansion, $\Theta$, is defined by
\be \Theta \equiv \nabla_a u^a, \ee while \be a^a \equiv
u^b\nabla_b u^a \ee is the acceleration vector. A useful
related quantity is the integrated volume
expansion along $u^a,$ defined as \be \alpha \equiv {1\over 3}\int d\tau \,
\Theta \quad \quad (\Theta = 3 \dot \alpha  )
\label{alpha_def}. \ee Since $\Theta/3$ corresponds to the
local Hubble parameter, one sees that the quantity $\alpha$ can
be interpreted as the number of e-folds measured along the
world-line of a cosmological observer with four-velocity $u^a$.
Finally, it is always possible to decompose the total energy
momentum tensor of a system as \be T_{ab} =(\rho+P) u_a u_b  +
q_au_b+ u_aq_b+ g_{ab} P + \pi_{ab}, \label{EMT2} \ee where
$\rho$, $P$, $q_a$ and $\pi_{ab}$ are respectively the energy
density, pressure,  momentum  and anisotropic stress tensor
measured in the frame defined by $u^a$.

\smallskip

The above quantities enter into the definition of the
generalized curvature perturbation on uniform energy density
hypersurfaces $\zeta_a$ (see
\cite{Langlois:2005ii,Langlois:2005qp}) \be \zeta_a \equiv D_a
\alpha - \frac{\dot \alpha}{\dot \rho} D_a \rho. \label{zeta}
\ee $\zeta_a$ is the key quantity we want to track the
evolution of. In the large scale limit (for which we are using the notation $\approx$),
it satisfies the simple
first-order evolution equation \be \dot \zeta_a \approx
\frac{\Theta^2}{3 \dot \rho} \Gamma_a, \label{zeta_evolution}
\ee where \be \Gamma_a \equiv D_a P
 -\frac{\dot{P}}{\dot  \rho} D_a \rho
\label{Gamma_def} \ee is the nonlinear nonadiabatic pressure
perturbation. For a barotropic fluid, $\Gamma_a=0$ and
$\zeta_a$ is conserved on large scales.  The relation
(\ref{zeta_evolution}) can then be seen as a generalization of
the familiar conservation law for $\zeta$, the linear curvature
perturbation on uniform energy hypersurfaces
\cite{Bardeen:1983qw,Lyth:2004gb}. Finally, note a useful
property of $\z_a$: in its definition Eq.~(\ref{zeta}) one can
replace  the spatial gradients $D_a$ by partial or covariant
derivatives,
 \be \zeta_a = \nabla_a
\alpha - \frac{\dot \alpha}{\dot \rho} \nabla_a \rho\,. \ee The
same property of course applies to similar combinations of
spatial gradients, such as $\Gamma_a$.

\subsection{Two scalar fields}

In this subsection, we specify the above formalism to the situation where two scalar fields
minimally coupled to gravity fill the universe. The corresponding Lagrangian density is
\be
{\cal L} = - \frac{1}{2}
\partial_a \phi_\I \partial^a \phi^\I -  V(\phi_K) ,
\label{lagrangian}
\ee
where $V$ is the potential. We assume, for simplicity,
canonical kinetic terms. Here and in the following, summation
over the field indices ($I,J = 1,2 $)  will be implicit. The
energy-momentum tensor derived from this Lagrangian reads
\be
T_{ab}= \partial_a \phi_\I \partial_b \phi^I -\frac{1}{2} g_{ab}
\left( \partial_c \phi_\I
\partial^c \phi^I + 2 V \right). \label{EMT}
\ee
Starting from the energy-momentum tensor (\ref{EMT}) one finds
\bea
\rho &\equiv&
T_{ab}u^au^b= \frac{1}{2} \left(   \dot \phi_I\dot\phi^I + \Dc_a
\phi^I \Dc^a \phi_\I \right) +V , \label{rhotot}\\
P &\equiv& \frac{1}{3}h^{ac}T_{ab}h^b_{\, c}= \frac{1}{2}  \left(
\dot \phi_I\dot\phi^I - \frac{1}{3} \Dc_a \phi_\I
\Dc^a \phi^I \right) -V, \label{Ptot}\\
q_a &\equiv&
-u^bT_{bc}h^c_{\, a}=  - \dot \phi_\I \Dc_a \phi^\I, \label{q}\\
\pi_{ab} &\equiv&   h_a^{\, c}T_{cd} h^d_{\, b}-P h_{ab}= \Dc_a
\phi_\I \Dc_b \phi^I -\frac{1}{3} \hc_{ab} \Dc_c \phi_\I \Dc^c \phi^\I .
\label{as}
\eea

\smallskip

In the two-field case, it is convenient to introduce a particular
basis in field space in which various field dependent
quantities are decomposed into adiabatic and entropy
components. In the linear theory, this decomposition
 was first introduced in
\cite{Gordon:2000hv} for two fields. For the multi-field case,
it is discussed in \cite{GrootNibbelink:2000vx,Langlois:2008mn}
in the linear theory and in \cite{Rigopoulos:2005xx} in the
nonlinear context. In our case, the corresponding basis
consists of a unit
vector $e_\sigma^I$ defined in the direction of the velocity of
the two fields, and thus {\em tangent} to the trajectory in
field space, and of a  unit vector $e_s^I$ defined along  the
direction {\em orthogonal} to it (with respect to the field
space metric $G_{IJ}$, taken here to be trivial, $G_{IJ}=
\delta_{IJ}$, with which the field indices are raised and
lowered), namely
\be
e_\sigma^\I \equiv  \frac{1}{\sqrt{\dot\phi_1^2+\dot\phi_2^2}}
\left(\dot \phi_1, \dot \phi_2 \right),
\qquad e_s^\I \equiv
 \frac{1}{\sqrt{\dot\phi_1^2+\dot\phi_2^2}}
\left(- \dot \phi_2, \dot \phi_1 \right)
 \label{e1} .
\ee
To make the notation shorter it is  convenient
to introduce the  angle $\theta$ defined by
\be
\cos \theta \equiv \frac{\dot\phi_1}{\sqrt{\dot\phi_1^2+\dot\phi_2^2}}\, ,
\quad \quad \sin \theta \equiv
\frac{\dot\phi_2}{\sqrt{\dot\phi_1^2+\dot\phi_2^2}}\, ,
\label{theta}
\ee
so that
\be
e_\sigma^\I =
\left(\cos \theta,\sin \theta  \right),
\qquad e_s^\I =
\left(-\sin \theta , \cos \theta \right).
\ee
This angle, in contrast with the linear theory case where it is a
background quantity that depends only on time,  is here an
inhomogeneous  quantity which depends on  time and space. By
taking the time derivative of the basis vectors $e_\sigma^\I$ and
$e_s^\I$, we get
\be
\dot e_\sigma^\I =  \dot \theta e_s^\I  , \qquad
\dot e_s^\I = - \dot \theta
e_\sigma^\I. \label{angle_1}
\ee
It is also convenient to introduce the {\em formal} notation
\be
\dot{\sigma} \equiv \sqrt{ G_{IJ} \dot{\p}^I\dot{\p}^J} = \sqrt{\dot\phi_1^2+\dot\phi_2^2}\,,
\ee
such that, for instance,
\be
e_\sigma^\I = \frac{\dot{\phi}^I}{\dot{\sigma}}.
\ee
Notice that generically,
the quantity $\sigmad$ {\em is not} the derivative along $u^a$ of a
scalar field $\sigma\,$; we just use it as short-hand notation.

\smallskip

Making use of the basis (\ref{e1}), one can then introduce  two
linear combinations of the scalar field gradients and thus define
two covectors, respectively denoted by $\sigma_a$ and $s_a$, as
\bea
\sigma_a    \equiv  e_{\sigma I} \nabla_a \phi^\I
\label{tan_ort1}, \\
 s_a \equiv e_{s I} \nabla_a \phi^\I
  \label{tan_ort2}.
\eea
These two covectors are called the {\em adiabatic} and {\em
entropy}  covectors, respectively, by analogy with the similar
definitions in the linear context \cite{Langlois:2008mn}.

\smallskip

Although one can derive a second-order (in time) equation for
the adiabatic covector (see \cite{Langlois:2006vv}), on large
scales it is simpler to resort to the evolution equation for
$\zeta_{a}$. In that purpose, it is useful to introduce the
covector \be \epsilon_a\equiv\Dc_a\rho- \frac{\dot \rho}{\dot
\sigma}\sp_a, \label{epsilon} \ee which can be interpreted as a
covariant generalization of the {\em comoving energy density}
perturbation. Indeed, it has been shown in
\cite{Langlois:2006vv} that if the shear can be neglected on
large scales, $\e_a$ is also negligible, in particular in
Eqs.~(\ref{zeta_ls}) and (\ref{s4}) below. On large scales, the
shear rapidly decreases in an expanding universe while it gets
far less blueshifted than the dominant energy contribution in
ekpyrotic models; therefore, one can safely neglect it in both
cases. However, to accurately describe the evolution of
perturbations in a local way on large scales, we will need to
keep track of it, as discussed below.

Then, in our two-field system, Eq.~(\ref{zeta_evolution})
reduces to \cite{Langlois:2006vv} \be \dot \zeta_a \approx \frac{2}{3 \dot \sigma^3}
V_{,\sigma}  \epsilon_a +\frac{2 \Theta}{ 3 \dot \sigma^2}
V_{,s} s_a  . \label{zeta_ls} \ee while the entropic equation
can be expressed as \be \ddot s_a +\Theta \dot s_a + \left(
V_{,ss}+3 \dot \theta^2 \right) s_a  \approx  - 2 \frac{\dot
\theta}{\dot \sigma} \epsilon_a . \label{s4} \ee In the above
equations, $V_{,\s} \equiv e_{\s}^I V_{,I}$, $V_{,s} \equiv
e_{s}^I V_{,I}$, $V_{,ss} \equiv e_{s}^I e_{s}^J V_{,IJ},$ etc. Recall that we are using the notation that
$\approx$ denotes an equality on large scales.
Eqs.~(\ref{zeta_ls}) and (\ref{s4}) are our two master
equations. They are exact on large scales and despite their
simple appearance, they contain all the nonlinearities of the
Einstein-scalar field equations on large scales. In the following
subsection, we will review how the introduction of a coordinate system
enables one to straightforwardly derive from them the linearized
and second-order perturbation equations, while we
derive new results at third order in the next section. Although
the route to pursue is straightforward, intermediate
calculations can become long. Therefore, for convenience, we
have collected in the appendix various background as well as
first- and second-order expressions that will be used in the
rest of this paper.

\subsection{Perturbation theory up to second order}

To relate the covariant approach with the more familiar
coordinate based formalism (for a recent review of cosmological
perturbations in this context, see \cite{Malik:2008im}), one
introduces generic coordinates $x^\mu=\{t, x^i\}$ to describe
an almost-FLRW spacetime. A prime will then denote a partial
derivative with respect to the cosmic  time $t$, i.e. ${}'
\equiv
\partial / \partial t$, since the dot has been reserved to
denote the Lie derivative with respect to $u^a$. Also, we
expand fields as follows: \be X(t,x^i) = \bar{X}(t) + \d
X^{(1)}(t,x^i) + \d X^{(2)}(t,x^i) + \d X^{(3)}(t,x^i),\ee {\it
i.e.} without factorial factors, where a quantity with an
over-bar is evaluated on the background. Here, $\d X^{(1)}$ refers to a quantity that solves the linearized equations of motion, $\d X^{(2)}$ the quadratic equations, and so on. Occasionally, and only
when we feel that the meaning is unambiguous, we drop the
superscript $^{(1)}$ for perturbations at linear order. Also,
in the following, $u^a$ is chosen such that $u_i=0$. The
remaining component $u_0$ is then determined solely in terms of
metric quantities.

\smallskip

By expanding Eqs.~(\ref{tan_ort1}) and (\ref{tan_ort2}) up to second order, one finds,
for $\sigma_i$ and $s_i$ respectively,
\bea
\delta\sigma_i=\partial_i\delta\sigma, \qquad
\d \s \equiv
\bar{e}_{\s I} \delta \phi^I \\
\label{dsigma}
\delta s_i=\partial_i \d s, \qquad
\d s \equiv \bar{e}_{s I} \delta \phi^I  ,
\eea
at linear order and
\bea
\label{sigi2} \delta \sigma_i^{(2)} &\equiv& \partial_i
\sis + \frac{\thetab'}{\sib'} \sif\partial_i \sf -\frac{1}{\sib'} V_i, \label{si_i} \\
\label{s_i}  \delta s_i^{(2)}  &\equiv& \partial_i \ss +
\frac{\sif}{\sib'}  \partial_i \sf',
\eea
at second order. We have used the definitions
\bea
\label{si2} \sis &\equiv&  \bar{e}_{\sigma I}  {\delta \phi^{I (2)}}  + \frac{1}{2 \sib'} \sf \sf',  \\
\ss &\equiv& \bar{e}_{s I}  {\delta \phi^{I (2)}}    -\frac{\sif}{\sib'}  \left( \sf' +
\frac{\bar \theta'}{2}
 \sif\right)
 \label{sis}
\eea
and introduced the
spatial vector
\be
V_i \equiv \frac{1}{2} (\sf \partial_i \sf' - \sf' \partial_i \sf), \label{V_def}
\ee
which we will discuss shortly.

The definitions of the entropy perturbations at first and
second order are chosen such that these quantities are
gauge-invariant. Given a vector $\xi^a$, the gauge
transformation it generates is defined by the transformation
law of tensors  \cite{Malik:2008im} (whose coordinate functions
are a particular case) \be
 {\bf \tilde T} \rightarrow e^{\L_{\xi}} {\bf T}\,.
\ee With the perturbative expansion $\xi=\sum_n \frac{1}{n!} \,
\xi_{(n)}$, the first and second-order perturbations of a
tensor  ${\bf T}$ are then found to transform as
\cite{Bruni:1996im} \be {\bf \delta T}^{(1)}\rightarrow {\bf
\delta T}^{(1)}+\L_{\xi_{(1)}} {\bf T}^{(0)},\quad {\bf \delta
T}^{(2)}\rightarrow {\bf \delta T}^{(2)}+\L_{\xi_{(2)}}{\bf
T}^{(0)}+\frac{1}{2}\L_{\xi_{(1)}}^2{\bf T}^{(0)}+
\L_{\xi_{(1)}} {\bf \delta T}^{(1)}.
\label{gauge_transformation} \ee Using these relations, it is
easy to verify that $\d s^{(1),(2)}$ are indeed gauge-invariant
on large scales.

The adiabatic perturbations are gauge-variant quantities, and
hence, their definition can be chosen for convenience. In fact,
it turns out that an inspection of the momentum density \bea
q_i= -\dot \s \s_i \label{qi} \eea typically provides a clue as
to what a convenient definition for $\d \s$ ought to be. A
straightforward expansion yields \be \delta q_i =-\partial_i
\left(\sib' \sif \right) \ee \bea \delta q_i^{(2)} =-\partial_i
\left(\sib' \sis +\frac{1}{2} \frac{\sib''}{\sib'} \sif^2 +
\thetab' \sif \sf \right) -\frac{1}{\sib'} \delta \epsilon
\partial_i \sif +V_i, \label{momentum_2_def} \eea and hence we
can see that the momentum density perturbations almost
automatically vanish when the adiabatic perturbations do. In
other words, setting the adiabatic perturbations to zero (as a
gauge choice) almost automatically corresponds to adopting
comoving gauge. The fly in the ointment is the spatial vector
$V_i$. However, as can be seen from its definition above, $V_i$
vanishes when $\sf'$ and $\sf$ have the same spatial
dependence, {\it i.e.} when $\sf'=f(t)\sf$. This condition is
fulfilled for super-Hubble modes in inflationary and also in
ekpyrotic models, as such relative spatial gradients are
heavily suppressed on large scales in both cases. Thus we can
define an approximate comoving gauge at second order, which
coincides with setting $\sif^{(1)}=\sis=0$.

\smallskip

Expanding now Eq.~(\ref{epsilon}), one finds
\be
\delta \epsilon_i = \partial_i \delta \epsilon, \qquad \delta
\epsilon \equiv \delta \r - \frac{\rb'}{\sib'} \d \s ,
\ee
where $\delta\epsilon$ is the first-order
comoving energy density perturbation, while at second order,
\be
\delta \epsilon_i^{(2)} = \partial_i \delta \epsilon^{(2)} +
\frac{\sif}{\sib'}
\partial_i \delta \epsilon^{(1)}{}' -3H\, V_i, \label{epsilon_i2}
\ee
with $\delta \epsilon^{(2)}$ defined by
\be
\delta \epsilon^{(2)} \equiv \rhos - {\rb'\over \sib'} \sis
 - \frac{\sif}{\sib'} \left[ \delta \epsilon^{(1)}{}'  +\frac{1}{2}
 {\left({\rb'\over \sib'}\right)}'
 \sif  +    \frac{\rb'}{\sib'} \bar \theta' \sf \right]
 . \label{delta_epsilon_def_2}
\ee
representing the energy density in the approximate comoving gauge on large scales.

\smallskip

As for the curvature perturbation on
uniform energy density hypersurfaces, it reads
\be
\delta \zeta_i =\partial_i \zeta, \qquad \zeta \equiv \d \a
-\frac{H}{\rb'} \d \r. \label{zeta_first}
\ee
\be
\zeta^{(2)}_i=\partial_i\zeta^{(2)}+\frac{\d \r}{\rb'}
\partial_i\zeta^{(1)}{}'
\label{delta_zeta_2}
\ee
with
\be
\zeta^{(2)}\equiv \d \a^{(2)} -\frac{H}{\rb'}\d \r^{(2)}
 - \frac{\d \r}{\rb'} \left[ \zeta^{(1)}{}' +
\frac{1}{2}\left(\frac{H}{\rb'}\right)'  \d \r \right] , \ee
Using the transformations (\ref{gauge_transformation}), one can
readily verify that $\d \epsilon^{(1),(2)}$ and
$\zeta^{(1),(2)}$ are gauge-invariant quantities on large
scales.

After these identifications of the desired
gauge-invariant quantities, the expansion of the master
equations Eqs.~(\ref{zeta_ls}) and (\ref{s4}) on large scales
gives (neglecting the first- and second-order comoving energy
density $\d \e$ and $\d \e^{(2)}$ at the end of the calculation) \bea \zeta'  \approx -
\frac{2H}{\sib'}\,\tb'  \delta s\, ,
\label{zeta_l} \\
\sf'' +3H\sf' +( \bar V_{,ss}+3 \thetab'{}^2 )
\sf \approx 0\,
\label{s_evol_1}
\eea
at first order, and
\bea
\zeta^{(2)}{}'&\approx& -\frac{H}{\sib'^2} \left[ 2\bar \theta'
\sib' \ss
 -  \left( \bar V_{,ss} + 4 \bar \theta'{}^2
\right)  \sf^2
+ \frac{\bar V_{,\sigma}}{\bar \sigma'} \delta s \delta s'
 \right] \, ,
   \label{zeta2_evol}   \\
\ss{}''+3H \ss{}'+\left(\bar V_{,ss}+3{\bar {\theta}}^{\prime
2}\right)  \ss &\approx&   -\frac{\bar \theta'}{\sib'}  \sf'{}^2
 - \frac{2}{\sib'}\left( \bar \theta''+ \bar \theta'
\frac{\bar V_{,\sigma}}{\sib'} -  \frac{3}{2} H \thetab'\right)
\sf \sf' \nn \\ &&
 - \left( \frac{1}{2} \bar V_{,sss} - 5\frac{\bar
\theta'}{\sib'} \bar V_{,ss} - 9 \frac{\bar \theta'{}^3}{\sib'}
\right)\sf^2 \label{s_evol_2} \eea at second order. It is
worth commenting that if one neglects the covector $\e_a$ in
Eqs.~(\ref{zeta_ls}) and (\ref{s4}) from the beginning, one can
write the second order equations of motion derived from them as
a total gradient, and hence in a local form as in
Eqs.~(\ref{zeta2_evol}) and (\ref{s_evol_2}), only if one
assumes throughout that $V_i=0$. If $\e_a$ is kept however, the equations
of motion become local, and the two limits of the equations
agree if the scalar quantities $\d \e$ and $\d \e^{(2)}$ are
neglected, as is appropriate.

These equations form a closed system: on large scales, the
entropy perturbation evolves independently of the adiabatic
component, and it sources the curvature perturbation. In the next
section, we will derive the corresponding third-order equations,
which are needed for the study of the primordial trispectrum of
curvature perturbation.

\section{Deriving the third order equations of motion}
\label{sectionThirdOrder}

Let us start by identifying the third order curvature perturbation on
a hypersurface of uniform energy density. By
expanding Eq.~(\ref{zeta}), one finds, after some
manipulations, that one can write \bea \delta \z_i^{(3)}
&\equiv& \partial_i  \z^{(3)} +\frac{\delta \r}{\rb'}\partial_i
\z^{(2)'} +\frac{\delta \r^{(2)}}{\rb'}
\partial_i \z^{(1)'} -\frac{\rb''}{2\rb'^3}\delta
\r^2\partial_i \z^{(1)'} + \frac{\delta
\r^2}{2\rb'^2}\partial_i \z^{(1)''},
\label{zetai3}
 \eea
with \bea \z^{(3)}&=&\delta \a^{(3)}-\frac{H}{\rb'}\delta
\r^{(3)}-\frac{1}{\rb'}\left(\delta \a'- \frac{H}{\rb'}\delta
\r'\right)\left(\delta \r^{(2)}-\frac{\delta \r \,\delta
\r'}{\rb'}+\frac{\rb'' \delta \r^2}{2 \rb^{'2}} \right)
-\frac{\delta \r}{\rb'} \left( \delta \a^{(2)'}-
\frac{H}{\rb'}\delta \r^{(2)'} \right)\nn \\ && +\frac{\delta
\r^2}{2 \rb^{'2}} \left(\delta \a''- \frac{H}{\rb'}\delta \r''
\right)+\frac{1}{\rb'}\left(\frac{H}{\rb'} \right)' \delta
\r^{(2)} \delta \r-\frac{1}{\rb^{'2}}\left(\frac{H}{\rb'}
\right)' \delta \r^2 \delta \r'-\frac{1}{6 \rb'} \left[
\frac{1}{\rb'} \left(\frac{H}{\rb'} \right)'\right]'\delta
\r^3. \label{zeta3-explicit} \eea Moreover, by applying the
gauge transformation at third order \cite{Bruni:1996im} \be
{\bf \delta T}^{(3)} \rightarrow {\bf \delta T}^{(3)}
+\L_{\xi_{(1)}} {\bf  \d T}^{(2)} +(\frac{1}{2}
\L^2_{\x_{(1)}}+\L_{\x_{(2)}}){\bf \delta T}^{(1)} +
(\L_{\x_{(3)}}+\L_{\x_{(1)}}\L_{\x_{(2)}}+\frac{1}{6}\L^3_{\x_{(1)}})
{\bf T}^{(0)} \ee to the covector $\zeta_a$ (for which the
formula simplifies since $\zeta_a^{(0)}=0$), one verifies that
$\z^{(3)}$ is indeed gauge invariant at third order on large
scales. From its definition Eq.~(\ref{zeta3-explicit}), one
sees that it has the desired interpretation as the third-order
curvature perturbation (or perturbation in the integrated
expansion $\a$) on uniform energy density hypersurfaces. It
also matches the definition of the same quantity on large
scales given in \cite{Enqvist:2006fs}. Its evolution equation
can be found by expanding Eq.~(\ref{zeta_evolution}): \bea
\z^{(3)'} &=&
-\frac{H}{\sb^{'2}}\Gamma^{(3)}-\frac{4}{3\sb^{'2}}\Gamma^{(2)}\z^{(1)'}
-\frac{1}{3H\sb^{'2}}\Gamma^{(1)}(\z^{(1)'})^2-\frac{2}{3\sb^{'2}}\Gamma^{(1)}\z^{(2)'}
\label{zeta3'} \eea where $\Gamma^{(3)}$ is defined in the same
way as $\zeta^{(3)}$ in Eqs.~(\ref{zetai3}) and
(\ref{zeta3-explicit}) with the replacements $\a \to P, \,
\zeta \to \Gamma$.

\bigskip

As the non-adiabatic pressure perturbation $\Gamma_a$ is
ultimately related to the presence of the entropic perturbation
(see Eq.~(\ref{zeta_ls})), one needs to identify the latter at
third order. This is similar to the analysis carried out for
the curvature perturbation, except that has to bear in mind
that $\d \s^{(2)}$ does not transform as the second-order
perturbation of a scalar quantity. Instead \bea \delta \s^{(2)}
\to \delta \s^{(2)} +\x_{(2)}^0 \sb' +\half \x_{(1)}^0
\left(\x_{(1)}^0 \sb'  \right)'+\x_{(1)}^0 \delta
\s'-\x_{(1)}^0 \tb' \delta s \eea where the last term is
different from the transformation properties of say $\delta
\r^{(2)}$. Then, using Eqs.~(\ref{deltaes}) and
(\ref{delta_es_2}), one finds \bea \delta s_i^{(3)} &\equiv&
\partial_i \delta s^{(3)} +\frac{\delta \s}{\sb'}\partial_i
\delta s^{(2)'} +\frac{\delta \s^{(2)}}{\sb'}
\partial_i \delta s' -\frac{\s''}{2\sb'^3}\delta
\s^2\partial_i \delta s' + \frac{(\delta
\s)^2}{2\sb'^2}\partial_i \delta s'' \nn \\ && +\frac{(\d s' +
2 \tb' \d \s)}{2 \sb^{'2}} \delta s \partial_i \delta s', \eea
where $\d s^{(3)}$ is a gauge invariant quantity at third order
on large scales, with the explicit definition \bea \delta
s^{(3)}&\equiv& \bar{e}_{s I} \delta \phi^{I (3)} -\frac{\delta
\s^{(2)}}{\sb'}\left(\delta s'+\tb' \delta \s
\right)-\frac{\delta \s}{\sb'}\delta s^{(2)'}-\frac{\delta
\s^2}{2 \sb^{'2}}  (\d s'' - \frac{\sb''}{\sb} \d s' +\tb^{'2}
\delta s)  -\frac{\delta \s^3}{6 \sb'}\left(\frac{\tb'}{\sb'}
\right)' \nn \\ && -\frac{\tb'}{2 \sb^{'2}}\delta s \delta s'
\delta \s. \label{s3} \eea Of course, one could have added
additional gauge-invariant terms to the definition of $\d
s^{(3)}$, {\it e.g.} terms involving cubic combinations of the
entropy perturbation at first order. Our definition is the most
natural as it represents the purely third order perturbations
of the fields perpendicular to the trajectory in the
approximate comoving gauge where $\sif^{(1)}=\sis=0$. One can
also build up $\d s^{(3)}$ by identifying the gauge-invariant
quantity that precisely reduces to $\bar{e}_{s I} \delta
\phi^{I (3)}$ in the approximate comoving gauge. Using the
transformation rules for the field perturbations, this leads to
the same definition Eq.~(\ref{s3}).

\smallskip

There is more freedom for the adiabatic perturbation as one is
not guided by gauge-invariance requirements. Expanding
Eq.~(\ref{tan_ort1}) using Eqs.~(\ref{deltaesigma}) and
(\ref{delta2esigma}), one obtains \bea \delta \s_i^{(3)}&=&
\partial_i \d\s^{(3)} + \frac{\tb'}{\sb'}\left( \d \s
\partial_i \d s^{(2)}+\d \s^{(2)} \partial_i \d s \right) +
\frac{\d \s}{\sb^{'2}}(\d s'+\tb'\d\s)\pt_i\d s' +
\frac{1}{2\sb^{'2}}(\d s'+\tb' \d \s)^2\pt_i \d\s\ \nn
\\ &&+ \left(\frac{\tb^{'2}}{\sb^{'2}}\d s + \left(\frac{\tb'}{\sb'}\right)'\frac{\d \s}{2}+\frac{1}{\sb'}\d s''-\frac{\sb''}{\sb^{'2}}\d s'\right)\d \s \pt_i\d s
-\frac{1}{\sb'}V_i^{(3)}-\frac{\tb'}{3\sb^{'2}}\d s V_i \,,
\label{sigmai3} \eea with \be \d\s^{(3)} \equiv \bar e_{\s I}
\d \p^{I (3)} +\frac{1}{2\sb'}(\d s \d s^{(2)'} + \d s^{(2)} \d
s') +\frac{\tb'}{6\sb^{'2}}\d s^2\d s' \label{delta_sigma_3} \ee
and \be V_i^{(3)} = \frac{1}{2}\left(\d s^{(2)} \pt_i \d s' +\d
s \pt_i \d s^{(2)'} - \d s^{(2)'} \pt_i \d s- \d s' \pt_i \d
s^{(2)}\right)\,, \ee which is the natural generalization to
third order of the non-local term $V_i$. The global consistency
of the formalism leads us to the particular choice
Eq.~(\ref{delta_sigma_3}) for $\d \s^{(3)}$ as will be clear
from the subsequent discussion of the momentum.

\smallskip

Expanding Eq.~(\ref{qi}), one finds, after a long
calculation, \bea q_i^{(3)}&=&-\partial_i \left(\sb'  \d
\s^{(3)}+\tb' \d \s \d s^{(2)} + \tb' \d \s^{(2)} \d s
+\frac{\sb''}{\sb'} \d \s^{(2)} \d \s +\frac{\tb'}{\sb'}(\d
\s)^2 \d s'+\frac{1}{2 \sb'} \d \s (\d s')^2 \right. \cr &+&\left. \half
\left(\frac{\sb'' \tb'}{\sb^{'2}}+\frac{\tb''}{\sb'} \right)\d
s \d \s^2- \frac{(\bar
V_{,ss}+\tb^{'2})}{2 \sb'} \d s^2 \d \s +\frac{\bar V_{,\s}}{2 \sb^{'2}}\d s \d s' \d \s \right. \cr &+&\left. \left(-\frac{2\bar
V_{,\s \s}-\tb^{'2}}{2 \sb'}+\frac{1}{2 \sb^{'2}}\left(
\frac{\bar \rho'}{\sb'} \right)'-\frac{(\sb'')^2}{2
\sb^{'3}}+\frac{\tb^{'2}}{2 \sb'} \right)\frac{\d \s^3 }{3}
 \right) \cr &-&
\frac{\d \e^{(2)} \partial_i \d \s}{\sb'}- \frac{\d \e
\partial_i \d \s^{(2)}}{\sb'}-\frac{\d \s \d \e' \partial_i \d
\s}{\sb^{'2}}+\frac{\d \e^2 \partial_i \d \s}{2
\sb^{'3}}+  \frac{\sb''}{\sb^{'3}}\d \e \d \s
\partial_i \d \s \cr &-& \frac{3H \d \s V_i}{\sb'}+ \frac{
\tb'}{\sb^{'2}} \d \e \partial_i( \d s  \d \s) +\frac{\bar
V_{,\s s}}{\sb'}\d \s \d s \partial_i \d \s +  \frac{\bar V_{,s
s}}{2\sb'} \d s^2 \partial_i \d \s \cr &+&V_i^{(3)}+\frac{4
\tb'}{3 \sb'}\d s V_i    +\frac{\d \e}{\sb^{'2}}V_i \eea The
third-order momentum is similar in form to the second-order
one, Eq.~(\ref{momentum_2_def}). In particular, it also cannot
directly be written as a total gradient. Of particular
relevance are the last three terms. While the last two involve
the second-order quantity $V_i$, negligible in the approximate
comoving gauge at second order, there also appears its
third-order generalization $V_i^{(3)}$. $V_i^{(3)}$ again
vanishes when the total entropy perturbation factorizes in
terms of its time and spatial dependence, and we neglect it for
the same reason that we neglect $V_i$ at second order, namely
that such differences of spatial gradients are heavily
suppressed on large scales in both inflationary and ekpyrotic
models.  When both $V_i$ and $V_i^{(3)}$ are negligible, one
can define the comoving gauge at third order on large scales as
$\d \s=\d \s^{(2)}=\d \s^{(3)}=0$.

Using the definitions above, we can define the (approximate)
comoving energy density at third order. Requiring that this
should be a gauge invariant quantity at third order on large
scales, reducing to the third-order perturbation of the energy
density in the approximate comoving gauge, we are led to
define \bea \e^{(3)}  &\equiv &  \delta
\r^{(3)}-\frac{\rb'}{\sb'} \delta \s^{(3)} - \frac{1}{\sb'}
\left(\delta \r'- \frac{\rb'}{\sb'}\delta
\s'\right)\left(\delta \s^{(2)}-\frac{\delta \s \,\delta
\s'}{\sb'}+\frac{\sb'' \delta \s^2}{2 \sb^{'2}} \right)
-\frac{\delta \s}{\sb'} \left( \delta \r^{(2)'}-
\frac{\rb'}{\sb'}\delta \s^{(2)'} \right) \cr &+& \frac{\delta
\s^2}{2 \sb^{'2}} \left(\delta \r''- \frac{\rb'}{\sb'}\delta
\s'' \right)+\frac{1}{\sb'}\left(\frac{\rb'}{\sb'} \right)'
\delta \s^{(2)} \delta
\s-\frac{1}{\sb^{'2}}\left(\frac{\rb'}{\sb'} \right)' \delta
\s^2 \delta \s'-\frac{1}{6 \sb'} \left[ \frac{1}{\sb'}
\left(\frac{\rb'}{\sb'} \right)'\right]'\delta \s^3 \cr
&-&\frac{\rb' \tb'}{\sb^{'2}}(\d s \d \s^{(2)}+\d s^{(2)} \d
\s)-\frac{\tb'}{\sb^{'2}} \d \r' \d s \d \s+\frac{2 \rb'
\tb'}{\sb^{'3}}\d \s \d \s' \d s-\frac{\rb'\tb^{'2}}{6
\sb^{'3}}\d \s^3-\frac{\rb'}{2 \sb^{'3}} \d \s \d s'^2 \cr &+&
\frac{\rb'(\bar V_{,ss}+\tb^{'2})}{2 \sb^{'3}} \d s^2 \d
\s+\left(\frac{\rb'' \tb'}{\sb^{'3}}+\frac{\rb' \tb''}{2
\sb^{'3}}-\frac{3 \tb' \rb' \sb''}{2 \sb^{'4}} \right) \d s \d
\s^2-\frac{\bar V_{,\s}\rb'}{2 \sb^{'4}}\d s \d s' \d \s
-\frac{3H \tb'}{\sb^{'2}}\d \s \d \e \d s . \eea Expanding
Eq.~(\ref{epsilon}), with the experience of the calculation of
$\zeta_i^{(3)}$, one obtains, using Eqs.~(\ref{dsigma1}),
(\ref{dsigma2}), (\ref{rsi1}) and (\ref{rsi2}) \bea
 \e_i^{(3)} & = &\partial_i \e^{(3)}+ \frac{\delta
\s}{\sb'}\partial_i \e^{(2)'} +\frac{\delta \s^{(2)}}{\sb'}
\partial_i \e^{(1)'} -\frac{\sb''}{2\sb'^3}\delta
\s^2\partial_i \e^{(1)'} + \frac{\delta
\s^2}{2\sb'^2}\partial_i \e^{(1)''} \cr &+& \frac{\d
\s}{\sb'}\left[ 3 V_i \left(\half
\sb'^2+3H^2\right)+\frac{\tb' \d s \partial_i \d \e'}{\sb'}-
\frac{3H \tb'}{\sb'}\left(\d \e \partial_i \d s-\d s \partial_i
\d \e\right) \right] \cr &-&3H V_i^{(3)}-\frac{4H \tb'}{\sb'}\d
s V_i +\frac{V_i \d \e'}{\sb^{'2}} \eea where the two non-local
terms $V_i$ and $V_i^{(3)}$ naturally appear.

\bigskip

We now have all the necessary ingredients to derive the
third-order equations of motion on large scales. Using the first
and second-order results (which can be read off from
Eqs.~(\ref{zeta_l}) and (\ref{zeta2_evol}))
 \bea \Gamma^{(1)} &\approx& -2\bar V_{,s}\d
s\,,  \\
\Gamma^{(2)} &\approx& -2\bar V_{,s}\d s^{(2)}-\bar V_{,ss}\d
s^2+\frac{\bar V_{,\s}}{\sb'}\d s \d s'\,, \eea we deduce from the
equivalent of Eq.~(\ref{zetai3}) that \bea \Gamma^{(3)} &\approx&
-2\bar V_{,s}\d s^{(3)}-2\bar V_{,ss}\d s \d
s^{(2)}+\frac{\bar V_{,\s}}{\sb'}(\d s^{(2)} \d
s)'-\frac{1}{3}\bar V_{,sss}\d s^3 + \frac{\th' \bar V_{,\s}+3 \bar V_{,\s
s}}{3\sb'}\d s'\d s^2\,. \eea Hence, from Eq.~(\ref{zeta3'}),
we obtain \bea \z^{(3)'} &\approx& \frac{2H}{\sb^{'2}}
\left(\bar V_{,s} \d s^{(3)}-\frac{1}{2\sb'}\bar V_{,\s}(\d s \d
s^{(2)})'-\frac{\tb'}{6\sb^{'2}}\bar V_{,\s}\d s^2\d s'+\bar V_{,ss}\d s
\d s^{(2)} - \frac{1}{2\sb'}\bar V_{,s\s}\d s^2\d
s'+\frac{1}{6}\bar V_{,sss}\d s^3\right) \nn \\ &&+
\frac{8H}{\sb^{'4}} \left(\bar V_{,s}^2 \d s \d
s^{(2)}-\frac{1}{2\sb'}\bar V_{,s}\bar V_{,\s}\d s^2\d s' +
\frac{1}{2}\bar V_{,s} \bar V_{,ss}\d s^3\right) + \frac{8H}{\sb^{'6}}
\bar V_{,s}^3 \d s^3  \,.
 \label{zeta3'}
\eea This equation describes the evolution of the curvature perturbation on
large scales at third order, and how it is sourced by the
entropy perturbation up to third order. It is the third-order
counterpart of Eqs.~(\ref{zeta_l}) and (\ref{zeta2_evol}). Note
that, once the potential $V$ becomes irrelevant, $\zeta$ is
conserved on large scales. This is important, as the potential
indeed becomes irrelevant in the approach to the big crunch in
ekpyrotic models.

\smallskip

To complete the system of equations, one needs to derive the
second order (in time) equation of motion for $\d s^{(3)}$.
This is done by expanding Eq.~(\ref{s4}) up to third order,
with the result that \bea && \d s^{(3)''} + 3H\d s^{(3)'} +
(\bar V_{,ss} + 3\tb^{'2}) \d s^{(3)} + 2\frac{\tb'}{\sb'}\d
s^{(2)'} \d s' \nn \\ && +
\left(2\frac{\tb''}{\sb'}+2\frac{\tb'\bar
V_{,\s}}{\sb^{'2}}-3H\frac{\tb'}{\sb'}\right)(\d s^{(2)} \d s)'
+\left(\bar V_{,sss}-10\frac{\tb' \bar
V_{,ss}}{\sb'}-18\frac{\tb^{'3}}{\sb'}\right)\d s^{(2)} \d s
\nn \\ && +\frac{\bar V_{,\s}}{\sb^{'3}}\d s'^3
+\left(\frac{\bar V_{,\s\s}}{\sb^{'2}} +3\frac{\bar
V_{,\s}^2}{\sb^{'4}}+3H\frac{\bar
V_{,\s}}{\sb^{'3}}-2\frac{\bar
V_{,ss}}{\sb^{'2}}-6\frac{\tb^{'2}}{\sb^{'2}}\right) \d s'^2 \d
s \nn \\ &&
+\left(-10\frac{\tb'\tb''}{\sb^{'2}}-\frac{3}{2\sb'}\bar
V_{,ss\s}-5\frac{\bar V_{,\s}\bar V_{,ss}}{\sb^{'3}}
-7\frac{\tb^{'2}\bar V_{,\s}}{\sb^{'3}}-3H\frac{\bar
V_{,ss}}{\sb^{'2}}+14H\frac{\tb^{'2}}{\sb^{'2}}\right)\d s' \d
s^2 \nn \\ && +\left(\frac{1}{6}\bar
V_{,ssss}-\frac{7}{3}\frac{\tb'}{\sb'}\bar V_{,sss}+2\frac{\bar
V_{,ss}^2}{\sb^{'2}} +21\frac{\tb^{'2}\bar
V_{,ss}}{\sb^{'2}}+27\frac{\tb^{'4}}{\sb^{'2}}\right)\d s^3 =
0\,. \label{s3'} \eea Similarly to the second-order
calculation, one can derive the same equation by neglecting the
covector $\e_a$ in Eq.~(\ref{s4}) from the beginning, if at the
same time one assumes that both $V_i=0$ and $V_i^{(3)}=0$.
\smallskip

Eqs.~(\ref{zeta3'}) and (\ref{s3'}) are the key results of
this section. They describe the coupled third order
perturbations of a system of two scalar fields on large scales
without any approximations. In the following sections, we will
use them to evaluate the primordial trispectrum in various
ekpyrotic models of the universe. However, we emphasize that these equations are valid on
superhorizon scales whatever the dynamics of the scale factor of the universe,
so that they are equally well suited to the study of
cosmological perturbations generated during a period of
inflation.

\section{Non-linearity parameters at third order} \label{sectionNonlinearity}

The observable that is relevant for comparison with
observations is the curvature perturbation $\z,$ and hence this
is the variable that we will focus on. In particular, we wish
to evaluate the trispectrum, {\it i.e.} the Fourier transform
of the 4-point function of the curvature perturbation. We are
interested in calculating the classical non-linearity generated
on super-Hubble scales. This was shown to be a good
approximation for the bispectrum in \cite{Koyama:2007if}, and
consequently we expect this to be a good approximation for the
trispectrum also. In other words, we expect the non-linearity
on sub-Hubble scales to be subdominant compared to the
non-linear classical super-Hubble evolution.

We only need to consider the connected part of the 4-point
function, {\it i.e.} the part that is not captured by products
of 2-point functions. If a distribution is exactly
gaussian, all information is contained in the 2-point function.
In particular this means that when $n$ is odd, all $n$-point
functions are zero, and when $n$ is even they can be decomposed
into products of 2-point functions. Here we therefore consider the part
of the 4-point function which is not already captured by the
product of two 2-point functions. Thus we define \be \langle
\z_{k_1}\z_{k_2}\z_{k_3}\z_{k_4}\rangle_c = (2\pi)^3 \d^3({\bf
k}_1+{\bf k}_2+{\bf k}_3+{\bf k}_4)T(k_1,k_2,k_3,k_4), \ee
where $T$ describes two different shape functions parameterized
by the $k$-independent parameters $\tau_{NL}$ and $g_{NL},$ see
\cite{Byrnes:2006vq} for more details. These are defined via
\be T=\tau_{NL}[P(k_{13})P(k_3)P(k_4) + {\rm 11 \,
permutations}]+\frac{54}{25}g_{NL}[P(k_2)P(k_3)P(k_4)+{\rm 3 \,
permutations}],\ee where $P(k)$ is the power spectrum defined
by the 2-point function \be \langle \z_{k_1}\z_{k_2}\rangle =
(2\pi)^3 \d^3({\bf k}_1+{\bf k}_2)P(k_1), \ee and ${\bf k}_{ij}=
{\bf k}_i + {\bf k}_j$.

The parameters $\tau_{NL}$ and $g_{NL}$ can be related to an
expansion of the curvature perturbation in terms of its
gaussian component, as follows. The Bardeen space-space metric
perturbation $\Phi_H$ \cite{Bardeen:1980kt} is expanded as \be
\Phi_H = \Phi_L + f_{NL} \Phi_L^2 + g_{NL} \Phi_L^3, \ee where
$\Phi_L$ denotes the linear, gaussian part (up to an
unimportant zero mode such that $\langle \Phi_H \rangle=0$).
During the era of matter domination, this variable is related
to the curvature perturbation via $\z = \frac{5}{3} \Phi_H,$ so
that we obtain the expansion \be \z= \z_L + \frac{3}{5}f_{NL}
\z_L^2 + \frac{9}{25}g_{NL} \z_L^3, \ee with $\z_L$ being the
linear, gaussian part of $\z.$ $\tau_{NL}$ is directly related
to the square of $f_{NL},$ explicitly \be \tau_{NL} =
\frac{36}{25} f_{NL}^2.\ee

The strategy for calculating the non-linearity parameters is
straightforward: first we solve the equations of motion for the
entropy perturbation up to third order in perturbation theory.
This allows us to integrate the equation of motion for $\zeta$,
also at the first three orders in perturbation theory, and then
we obtain the non-linearity parameters by evaluating  \bea
f_{NL} &=& \frac{5}{3}
\frac{\int_{t_i}^{t_f}\z^{(2)'}}{(\int_{t_i}^{t_f}
\z^{(1)'})^2}
\\  g_{NL} &=& \frac{25}{9}
\frac{\int_{t_i}^{t_f}\z^{(3)'}}{(\int_{t_i}^{t_f}
\z^{(1)'})^3},\eea where the integrals are evaluated from the
time $t_i$ that the ekpyrotic phase begins until the conversion
phase has ended at $t_f$ and $\zeta$ has evolved to a constant
value. The next sections will deal with the details of these
calculations.

\section{Ekpyrotic examples} \label{sectionEkpyrotic}

In ekpyrotic and cyclic models of the universe, the standard
cosmological puzzles, such as the horizon and flatness
problems, are resolved by a phase of slow contraction before
the big bang \cite{Khoury:2001wf}. During this ekpyrotic phase,
the equation of state $w_{ek}$ (the ratio of pressure to energy
density) is very large, with the consequence that the ekpyrotic
matter (usually modeled as a scalar field with a steep and
negative potential) quickly comes to dominate the energy budget
in the universe. In other words, the contributions to the total
energy density due to spatial curvature and anisotropies, for
example, become suppressed, and in this way the flatness
problem is solved. The horizon problem does not even arise, as
there was plenty of time before the big bang for different
regions of the universe to be in causal contact with each
other.

Interestingly, and just as in inflation, the ekpyrotic phase
can (quantum-mechanically) produce a nearly scale-invariant
spectrum of density perturbations at the same time as
(classically) flattening the universe. However, early studies \cite{Khoury:2001zk,Lyth:2001pf,Creminelli:2004jg}
of single-field ekpyrotic models uncovered a subtlety related
to the fact that the perturbations are generated during a
contracting phase: namely, the mode that picks up a
scale-invariant spectrum corresponds to a small time-delay
perturbation to the big crunch, and this mode turns out to be
the decaying mode in a contracting universe. The adiabatic
mode, on the other hand, picks up a blue spectrum. Hence,
unless these two modes mix during the big crunch - big bang
transition (which they might do due to higher-dimensional
effects for example \cite{Tolley:2003nx,McFadden:2005mq}), the spectrum of the growing mode during
the subsequent expanding phase will be incompatible with
observations.

However, it is unnatural to restrict the analysis to a single
scalar field. In models inspired by heterotic M-theory for
example \cite{Horava:1995qa,Horava:1996ma,Lukas:1998yy,Lukas:1998tt}, there are two universal scalars, namely the radion
mode and the volume modulus of the internal Calabi-Yau
manifold. The radion determines the distance between two
orbifold planes, and in this braneworld scenario, our currently
observed universe is identified with one of the orbifold
planes. In this picture, the ekpyrotic potential corresponds to
an attractive force between the orbifold planes, which
eventually causes the branes to collide. This collision then
looks like a big bang to an observer on the brane. Quickly
afterwards, the distance between the branes becomes fixed
again, but the branes themselves expand during the phases of
radiation, matter and dark energy domination. Actually, the
dark energy can be modeled by the same inter-brane force that
is at play during the ekpyrotic phase, it just corresponds to a
rather flat plateau at large field values. Inexorably though,
the branes are attracted to each other again, a new ekpyrotic
phase takes place, which is then swiftly followed by a new
brane collision, etc. In this way a cyclic picture naturally
emerges \cite{Steinhardt:2001st}.

But back to the two scalar fields: the theory we are
considering can be described by the effective action
\cite{Lehners:2006ir} \be S=\int
\sqrt{-g}[R-\frac{1}{2}(\pt\phi_1)^2
-\frac{1}{2}(\pt\phi_2)^2-V(\phi_1,\phi_2)],\ee where $\phi_1$ and $\phi_2$ are related by a field redefinition to the radion and the Calabi-Yau volume modulus. We are assuming
that during the ekpyrotic phase, both fields feel an
ekpyrotic-type potential, {\it e.g.} \be V(\phi_1,\phi_2) =-V_1
e^{-c_1 \phi_1} - V_2 e^{-c_2 \phi_2}.
\label{potential2field}\ee In fact, as discussed in detail in
\cite{Buchbinder:2007tw}, it is much more natural to discuss
the evolution in terms of the variables $\s$ and $s$ pointing
transverse and perpendicular to the field velocity respectively
(these are the variables we have used in sections
\ref{sectionCovariant} and \ref{sectionThirdOrder}). In terms
of those variables, the above potential can be re-expressed as
an ekpyrotic potential for $\s$ combined with a tachyonic
potential in the transverse direction. The evolution proceeds
along this unstable ridge in the potential. This instability is
necessary for the amplification of scale-invariant entropy
perturbations \cite{Tolley:2007nq}, and leads to an interesting
global structure \cite{Lehners:2008qe}.

Recasting the potential in terms of $\s$ and $s$ in fact
suggests that, in the absence of a microphysical derivation, we
should use a more general, parameterized, form for the
potential instead of (\ref{potential2field}). The feature that
we want to keep is that we should still obtain a nearly
scale-invariant spectrum of entropy perturbations. Hence,
during the ekpyrotic phase, we require the potential to have
the following features: 1. Along the $\s$ direction, it must be
of ekpyrotic form, {\it i.e.} negative and steep; more
precisely, we want $\bar{V}_{,\s\s} \approx -2/t^2.$ 2. Along
the $s$ direction, it must be tachyonic, {\it i.e.} we need
$\bar{V}_{,s} = 0$ and $\bar{V}_{,ss}<0$. As we just said, this
instability is necessary in order for entropy perturbations to
grow. 3. In order to obtain a nearly scale-invariant spectrum
of entropy fluctuations, we must have $\bar{V}_{,\s\s} \approx
\bar{V}_{,ss}.$ Small deviations from these relations lead to
small differences for the tilt of the spectrum of fluctuations
\cite{Buchbinder:2007tw}. Thus we write the potential as \be
V_{ek}=-V_0 e^{\sqrt{2\e}\s}[1+\e s^2+\frac{\k_3}{3!}\e^{3/2}
s^3+\frac{\k_4}{4!}\e^2 s^4+\cdots],
\label{potentialParameterized}\ee where we expect $\k_3,\k_4
\sim {\cal O}(1)$ and where $\e \sim {\cal O}(10^{2})$ is
related to the equation of state $w_{ek}$ via
$\e=3(1+w_{ek})/2.$ For exact exponentials of the form (\ref{potential2field}), one has $\k_3=2\sqrt{2}(c_1^2-c_2^2)/|c_1 c_2|$ and $\k_4=4(c_1^6 + c_2^6)/(c_1^2 c_2^2(c_1^2 + c_2^2)).$
Before continuing, we should comment on a
subtlety: strictly speaking, $s$ is not a field, since by
definition $s=0$ along the background trajectory. Hence, when
we expand the potential in terms of $s$, we simply use $s$ to
denote the direction transverse to the ridge of the potential.
We can only use the more compact notation here because the
background trajectory is a straight line during the ekpyrotic
phase, and hence the meaning of $s$ is unambiguous: in terms of
the original fields, one could use the replacement
$s=(c_2\phi_2-c_1\phi_1)/\sqrt{c_1^2+c_2^2},$ and this would
give identical results.

Along the ridge, the background scaling solution is given by
\be a(t)=(-t)^{1/\e} \qquad \bar\s=-\sqrt{\frac{2}{\e}}\ln
\left(-\sqrt{\e V_0} t\right) \qquad \bar{s}=0,
\label{ScalingSolution}\ee where time runs from large negative
values at the beginning of the ekpyrotic phase to $t=0$ at the
big crunch. This scaling solution corresponds to a straight
line trajectory in scalar field space. At the same time, nearly
scale-invariant entropy perturbations are produced and
amplified, their final amplitude being determined by the depth
of the potential. Subsequently, any bending of the trajectory
in field space will result in the conversion of entropy
perturbations into curvature (adiabatic) perturbations, which
inherit the same spectral index \cite{Lehners:2007ac};
accordingly, this mechanism is termed the entropic mechanism.

A bending of the trajectory is expected in many models on
generic grounds - we will discuss two cases in detail in
sections \ref{subsectionKineticConversion} and
\ref{subsectionEkpyroticConversion}. The first is where the
bending occurs during the kinetic energy dominated phase that
leads up to the big crunch (``kinetic conversion''), and the
second is where the trajectory starts out slightly off the
ridge, so that after some time, but still during the ekpyrotic
phase, it naturally rolls off the ridge to one side
(``ekpyrotic conversion''). As we will see, both the sign and
the amplitude of the non-linearity parameters depend rather
crucially on the details of this conversion process.

Before continuing, we should state an important assumption that
we are making in this work: namely, we assume that the
curvature perturbation passes through the big crunch - big bang
transition essentially unchanged. The reason for doing so is
that the perturbations we are considering are vastly larger
than the horizon size around the time of the crunch, and hence, due
to causality, it seems reasonable to assume that
long-wavelength modes suffer no change. Nevertheless, in the
absence of a complete quantum-gravitational treatment of the
big crunch - big bang transition, this statement remains an
assumption subject to possible revision in the future.

\subsection{Initial conditions for the entropy mode}

During the ekpyrotic phase, the trajectory is a straight line
with $\tb'=0,$ (see Eqs.~(\ref{theta}) and (\ref{angle_1})) and so the equation of motion for $\d s^{(3)}$
simplifies to
\bea
&& \d s^{(3)''} + 3H\d s^{(3)'} + \bar V_{,ss} \d s^{(3)} +\bar V_{,sss}\d s^{(2)} \d s \nn \\ && +\frac{\bar V_{,\s}}{\sb^{'3}}\d s'^3 +\left(\frac{\bar V_{,\s\s}}{\sb^{'2}}
+3\frac{\bar V_{,\s}^2}{\sb^{'4}}+3H\frac{\bar V_{,\s}}{\sb^{'3}}-2\frac{\bar V_{,ss}}{\sb^{'2}}\right) \d s'^2 \d s \nn \\ && +\left(-\frac{3}{2\sb'}\bar V_{,ss\s}-5\frac{\bar V_{,\s}\bar V_{,ss}}{\sb^{'3}} -3H\frac{\bar V_{,ss}}{\sb^{'2}}\right)\d s' \d s^2 +\left(\frac{1}{6}\bar V_{,ssss}+2\frac{\bar V_{,ss}^2}{\sb^{'2}} \right)\d s^3 = 0 \,. \eea
Using the following useful expressions, valid during the ekpyrotic phase,
\bea
&& \sb' = -\frac{\sqrt{2}}{\sqrt{\e}t} \qquad \bar{V}= -\frac{1}{\e t^2} \nn \\
&& \bar{V}_{,\s}= -\frac{\sqrt{2}}{\sqrt{\e}t^2} \qquad
\bar{V}_{,\s\s}= -\frac{2}{t^2} \qquad \bar{V}_{,s\s} = 0
\qquad
\bar{V}_{,ss\s} = -\frac{2\sqrt{2\e}}{t^2} \nn \\
&& \bar{V}_{,s}= 0 \qquad \bar{V}_{,ss} = -\frac{2}{t^2} \qquad
\bar{V}_{,sss}= -\frac{\k_3 \sqrt{\e}}{t^2}    \qquad
\bar{V}_{,ssss}=- \frac{\k_4 \e}{t^2},  \nn \eea together with
the first- and second-order entropic equations
(\ref{s_evol_1}), (\ref{s_evol_2}) in the limit when $\tb'=0$,
we iteratively find that the entropy perturbation (to leading
order in $1/\e$) can be expanded as \be \d s= \d s_L +
\frac{\k_3 \sqrt{\e}}{8}\d s_L^2 +
\e(\frac{\k_4}{60}+\frac{\k_3^2}{80}-\frac{2}{5})\d s_L^3,
\label{entropyInitialCondition}\ee with $\d s_L \propto 1/t.$
The above equation specifies the initial conditions for the
start of the conversion phase.

\subsection{Integrated contributions during the ekpyrotic phase}

There are contributions to both $f_{NL}$ and $g_{NL}$ even
during the ekpyrotic phase, since some of the terms in the
equation for $\z'$ do not vanish when $\tb'=0.$ Specifically,
during the ekpyrotic phase, we have ($\z$ here is the full
$\z^{(1)}+\z^{(2)}+\z^{(3)}$) \bea \z' &=&
\frac{H}{\sb^{'2}}[\bar V_{,ss}\d s^2 -\frac{\bar
V_{,\s}}{\sb'}\d s \d s' +\frac{\bar V_{,sss}}{3}\d s^3] \nn \\
&\approx& -\frac{1}{t}\d s^2-\frac{1}{2}\d s\d s'-\frac{\k_3
\sqrt{\e}}{6t}\d s^3 \nn
\\ &\approx& -\frac{\d s_L^2}{2t}- \frac{11\k_3\sqrt{\e}\d
s_L^3}{48t}\eea This leads to \bea f_{NLintegrated} &=&
\frac{5}{12}\frac{[\d
s_L(t_{ek-end})]^2}{[\zeta_L(t_{f})]^2} \\
g_{NLintegrated} &=& \frac{275}{1296}\k_3\sqrt{\e}\frac{[\d
s_L(t_{ek-end})]^3}{[\z_L(t_{f})]^3},\eea where
$t_{f}$ denotes the time at the end of the conversion
stage, when $\z$ becomes constant on large scales, and $t_{ek-end}$ denotes the time at the end of the ekpyrotic phase.

Numerically speaking, both $f_{NLintegrated}$ and $g_{NLintegrated}$ are typically
subdominant contributions to the total non-linearity
parameters.

\begin{figure}[t]
\begin{center}
\includegraphics[width=0.7\textwidth]{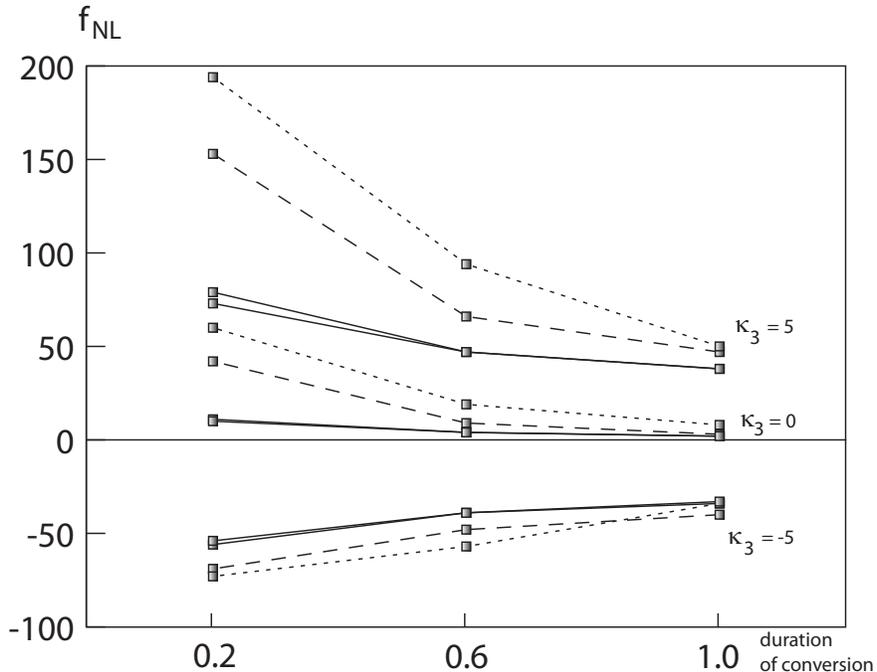}
\caption{\label{FigurefNL} {\small $f_{NL}$ as a function of the duration of conversion and for the values
$\k_3=-5,0,5$ and $\e=36$. In each case, we have plotted the results for four different reflection potentials, with the simplest potentials ($\phi_2^{-2},(\sinh{\phi_2})^{-2}$) indicated by solid lines, while the dashed ($(\sinh{\phi_2})^{-2}+(\sinh{\phi_2})^{-4}$) and dotted ($\phi_2^{-2}+\phi_2^{-6}$) lines give an indication of the range of values that can be expected. As the conversions become smoother, the
predicted range of values narrows, and smooth conversions lead
to a natural range of about $-50 \lesssim f_{NL} \lesssim + 60$ or so.
}}
\end{center}
\end{figure}

\begin{figure}[t]
\begin{center}
\includegraphics[width=0.7\textwidth]{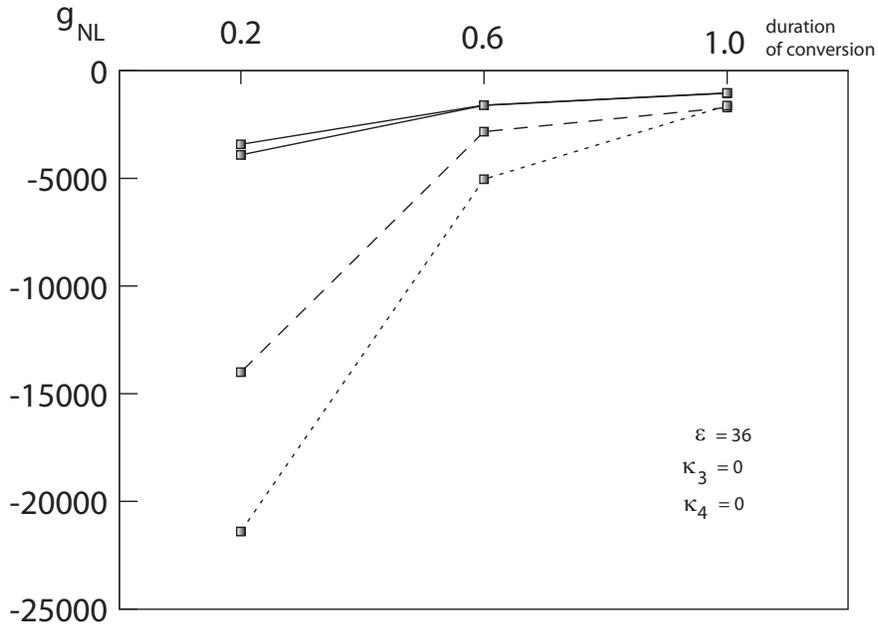}
\caption{\label{Figure1} {\small $g_{NL}$ as a function of the duration of conversion, with $\k_3=\k_4=0$ and for four different reflection potentials, with the same line style assignments as in Fig. \ref{FigurefNL}. As the conversions become smoother, the predicted range of values narrows considerably, allowing us to make rather definite predictions.
}}
\end{center}
\end{figure}

\begin{figure}[t]
\begin{center}
\includegraphics[width=0.7\textwidth]{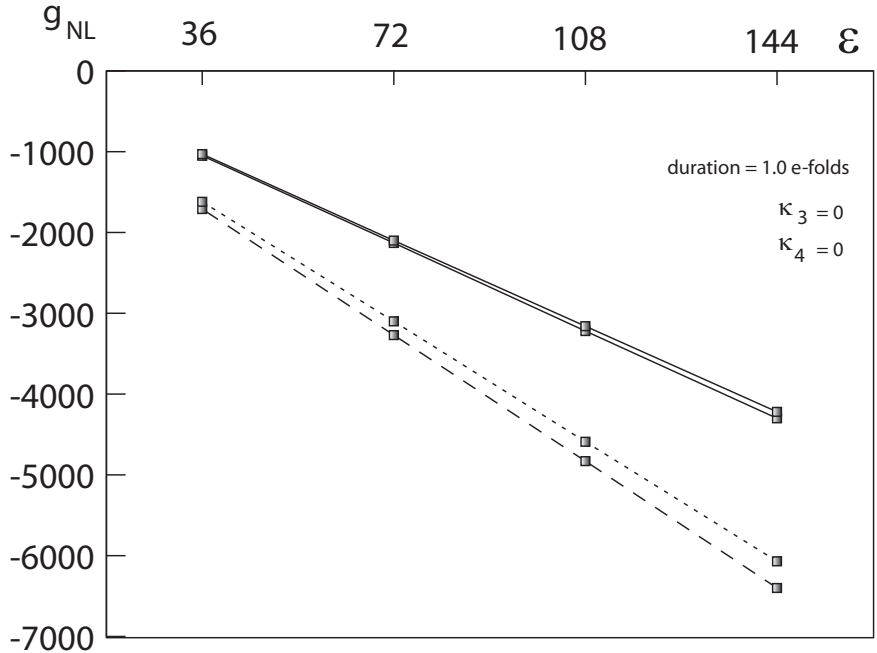}
\caption{\label{Figure2} {\small This figure shows $g_{NL}$ to be proportional to $\e$, {\it i.e.} proportional to the equation of state $w_{ek}.$
}}
\end{center}
\end{figure}

\begin{figure}[t]
\begin{center}
\includegraphics[width=0.7\textwidth]{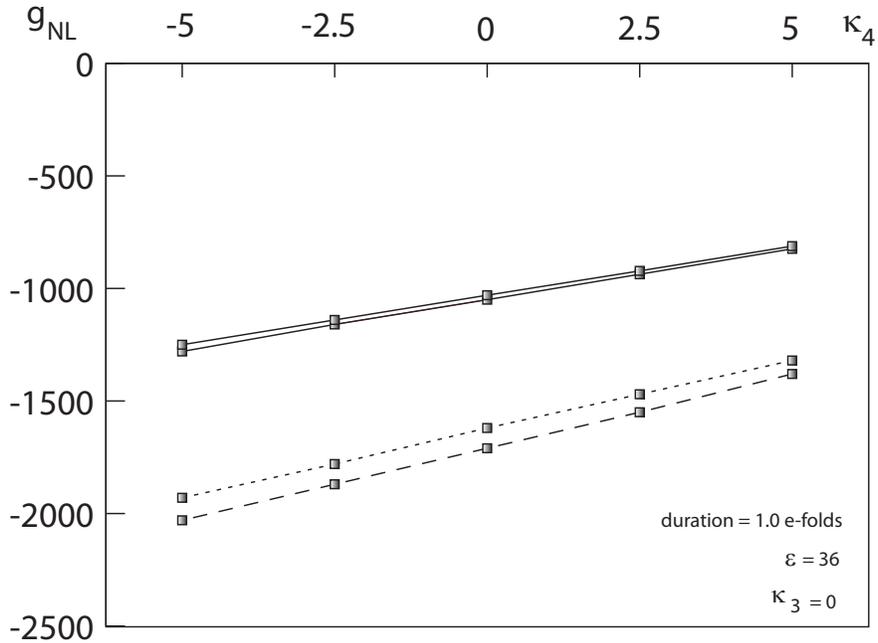}
\caption{\label{Figure3} {\small This figure indicates that $g_{NL}$ depends linearly in $\k_4,$ the parameter we are using to specify the fourth derivative of the ekpyrotic potential with respect to the entropic direction.
}}
\end{center}
\end{figure}

\begin{figure}[t]
\begin{center}
\includegraphics[width=0.7\textwidth]{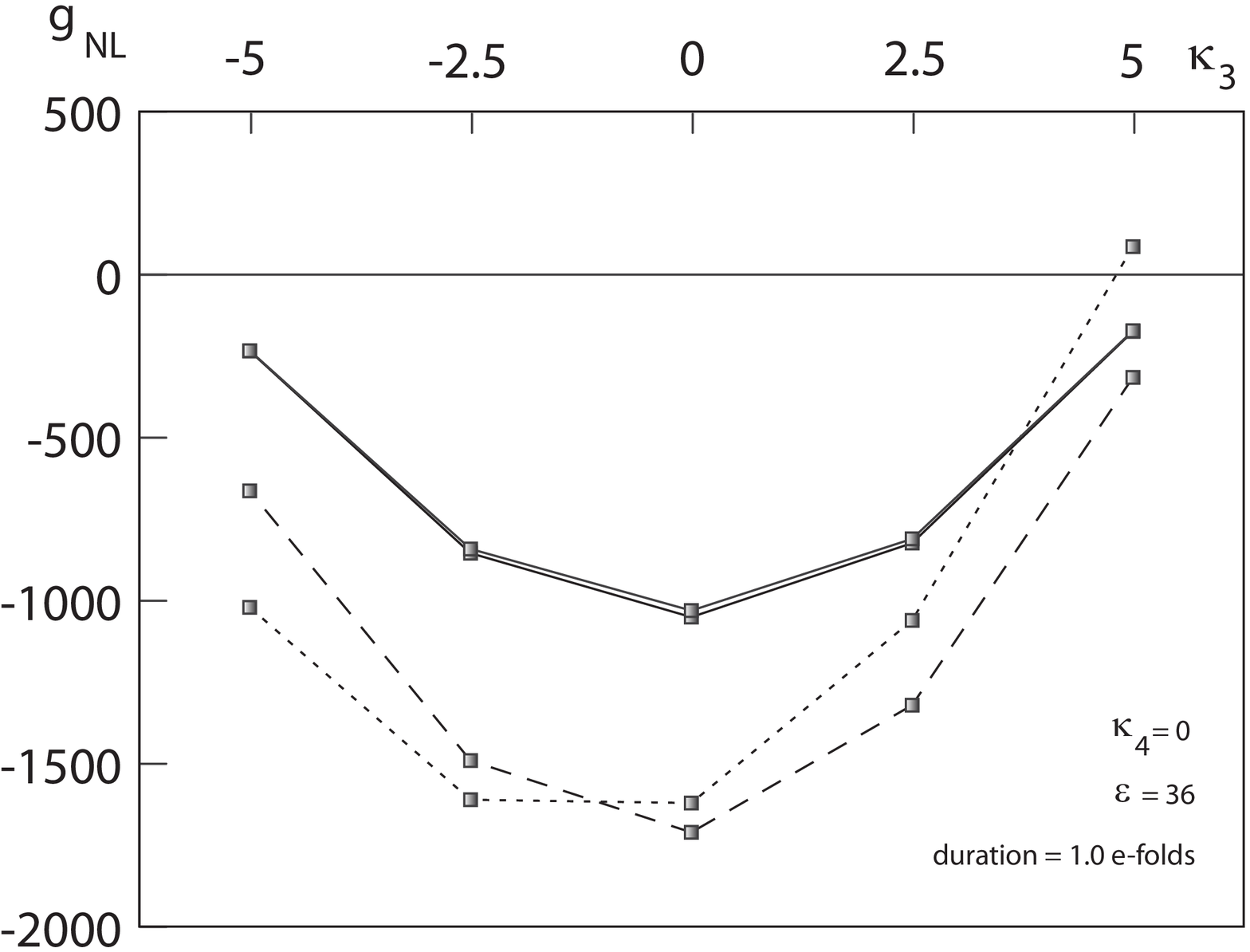}
\caption{\label{Figure4} {\small $g_{NL}$ can be seen to depend approximately quadratically on $\k_3,$ the parameter determining the third derivative of the ekpyrotic potential with respect to the entropic direction.
}}
\end{center}
\end{figure}

\subsection{Kinetic conversion} \label{subsectionKineticConversion}

In the original ekpyrotic and cyclic models, the phase
dominated by the steep, ekpyrotic potential $V_{ek}$ comes to
an end (at $t=t_{ek-end}<0$) before the big crunch/big bang
transition (at $t=0$), and as the ekpyrotic potential becomes
negligible the universe becomes dominated by the kinetic energy
of the scalar fields. In this subsection, we consider the case
where the conversion from entropic to curvature perturbations
occurs during this kinetic energy dominated phase. This occurs
naturally in the heterotic M-theory embedding of the cyclic
model \cite{Lehners:2006pu} (briefly discussed towards the
beginning of this section) because the negative-tension brane
bounces off a spacetime singularity (see \cite{Lehners:2007nb}
for a detailed discussion)  --  thus creating a bend in the
trajectory in field space in the 4-dimensional effective theory
\cite{Lehners:2006ir}-- before it collides with the
positive-tension brane (the latter collision corresponds to the
big crunch-big bang transition). This bending of the trajectory
automatically induces the conversion of entropy to curvature
perturbations \cite{Lehners:2007ac}. From the viewpoint of the
4-dimensional effective theory, the bounce of the
negative-tension brane corresponds to the trajectory in scalar
field space being reflected off a boundary of scalar field
space, and it can be modeled by having a repulsive potential
$V_{rep}$ in the vicinity of the boundary. Here we take the
boundary to be located at $\phi_2=0,$ and we only consider
conversions during which $\th'>0$ -- other cases can be related
to these by an appropriate change of coordinates. The repulsive
potential can in principle be calculated, given a specific
matter configuration on the negative-tension brane
\cite{Lehners:2007nb}. Here, in order to be general, we
consider four different forms for the repulsive potential that
we consider to be representative, namely \be V_{rep} \propto
\phi_2^{-2},\, \phi_2^{-2}+\phi_2^{-6},\,
(\sinh{\phi_2})^{-2},\,
(\sinh{\phi_2})^{-2}+(\sinh{\phi_2})^{-4},
\label{potentialRepulsive}\ee with the overall magnitude
adjusted in order to obtain various values for the duration of
the conversion (see below). These potential forms should give
an indication of the range of values that one can expect the
non-linearity parameters to take.

An important parameter turns out to be the duration of the
conversion, measured by the number of Hubble times during which most, say 90 percent, of
the conversion takes place, {\it i.e.} the duration is measured by the number of e-folds by which the scale factor shrinks during conversion. Conversions lasting less than about $0.2$ Hubble times
correspond to what we call sharp conversions, while those that last on the order of $1$ Hubble time correspond to smooth conversions. For
$f_{NL}$, the range of predicted values is considerably
narrower as the conversion becomes smoother
\cite{Lehners:2007wc,Lehners:2008my}, as illustrated in Fig.
\ref{FigurefNL}. It is possible to estimate $f_{NL}$
semi-analytically, with the result that for smooth conversions
\cite{Lehners:2008my}, \be f_{NL} \sim \frac{3}{2} \, \k_3
\sqrt{\e}+5. \label{FittingFormulafNL}\ee The values for
$\tau_{NL}=36/25 f_{NL}^2$ follow directly from these
calculations, and hence they do not require an extensive
discussion.

In order to calculate $g_{NL},$ we have solved and integrated
numerically the equations of motion (\ref{s3'}) and
(\ref{zeta3'}), using the expression
(\ref{entropyInitialCondition}) as the initial condition for
the entropy perturbation. $\zeta$ is initially zero. The
initial conditions for the background trajectory are taken to
be those suggested by the heterotic M-theory embedding
\cite{Lehners:2006ir}, {\it i.e.} we take the incident angle of
the trajectory with respect to the boundary surface $\phi_2=0$
(and thus the reflection potentials) to be $\pi/6$. On changing
this angle, we expect our results to change only modestly
\cite{Lehners:2008my}.

The results of our numerical calculations are shown in Figs.
\ref{Figure1} to \ref{Figure4}. In each case, we have plotted
the results obtained for the four repulsive potentials
(\ref{potentialRepulsive}). Fig. \ref{Figure1} shows that, even
more so than for the calculation of $f_{NL}$ the range of
predicted values for $g_{NL}$ narrows drastically as the
conversion process becomes smoother. In fact, for sharp
conversions, typical values are very large in magnitude, and we
expect these to be observationally disfavored shortly, if they
aren't ruled out already. Thus, phenomenologically speaking, it
is much more interesting to focus on smooth
conversions. In this respect, it is important to note that for
smooth conversions, the predicted range of values for $f_{NL}$
lies within the current observational bounds
\cite{Komatsu:2008hk,Smith:2009jr}. Hence, we will now focus on
conversions that occur while the scale factor evolves by more
than about half an e-fold.

The next three figures show the dependence of $g_{NL}$ on the
potential parameters $\k_3,\k_4$ and $\e=3(1+w_{ek})/2,$ as
defined in (\ref{potentialParameterized}). Fig. \ref{Figure2}
shows that $g_{NL}$ is proportional to $\e$, and we have
verified that this result holds independently of the values of
$\k_3$ and $\k_4.$ Similarly, Fig. \ref{Figure3} indicates that
$g_{NL}$ depends linearly on $\k_4$ (independently of $\e$ and
$\k_3$), while Fig. \ref{Figure4} indicates that $g_{NL}$
depends approximately quadratically on $\k_3$ (independently of
$\e$ and $\k_4$). In fact, all our numerical results indicate
that $g_{NL}$ scales with $\e,\k_3,\k_4$ exactly as the
third-order coefficient in the expression
(\ref{entropyInitialCondition}) for the entropy perturbation
during the ekpyrotic phase, and we have found that we can fit
all our data by the approximate formula \be g_{NL} \sim 100 \,
\e(\frac{\k_4}{60}+\frac{\k_3^2}{80}-\frac{2}{5}),
\label{FittingFormulagNL}\ee {\it i.e.} $g_{NL}$ is roughly
speaking directly proportional to the third-order coefficient
in the expansion (\ref{entropyInitialCondition}) of the entropy
perturbation during the ekpyrotic phase. This formula is one of
the key results of this paper. It shows that $g_{NL}$ is
proportional to the equation of state of ekpyrosis $w_{ek} \sim
\e.$ Hence it scales in the same way as $\tau_{NL} \propto
f_{NL}^2 \propto (\sqrt{\e})^2.$ Also, from the fitting formula
for $f_{NL}$ (\ref{FittingFormulafNL}), we can deduce that a
large $|f_{NL}|$ tends to make $g_{NL}$ more positive due to
its dependence on $\k_3$. Most significant is perhaps the term
independent of $\k_3,\k_4.$ This term is negative and typically
of order 1000, and it implies that $g_{NL}$ can be expected to
be negative unless $|\k_3|$ is large and/or $\k_4$ is large and
positive. In fact, the smaller $|f_{NL}|$ is, the more `tuned'
the value of $\k_4$ has to be in order to make $g_{NL}$
positive -- for $\k_3 \approx 0$, we would require $\k_4
\approx 25$ just in order to make $g_{NL}$ zero. Hence, any
accidental degeneracy at the level of $f_{NL}$ between simple
inflationary models (in which both $f_{NL}$ and $g_{NL}$ are
expected to be very close to zero) and cyclic models is
extremely likely to be broken at the level of the trispectrum.
Our result indicates that it is far from true that ``anything
goes'', {\it e.g.} it would be natural to find $f_{NL}=20$ and
$g_{NL}=-1500,$ while it would be very unnatural to get
$f_{NL}=20$ together with $g_{NL}=+3000.$ Thus, for the
currently best-motivated cyclic models in which the
perturbations are converted during the kinetic phase, Eq.
(\ref{FittingFormulagNL}) provides a distinct observational
imprint.

For completeness, we should note that there is the possibility
that the dominant amount of conversion could occur after the
big bang, via the mechanism of modulated preheating suggested
by Battefeld \cite{Battefeld:2007st}. The idea here is that
massive matter fields could be produced in large quantities at
the brane collision and could dominate the energy density right
after the big bang. If these massive fields were to couple to
ordinary matter via a function $h(\delta s)$, then their decay
into ordinary matter would occur at slightly different times
for different values of $\delta s.$ In this way, the ordinary
matter perturbations would inherit the entropic perturbation
spectrum. This conversion would typically also happen during a phase
of kinetic energy domination, so that we would expect the
orders of magnitude of $f_{NL}$ and $g_{NL}$ to be similar to
the examples above. However, in the absence of a detailed
model, which would determine $h(\d s),$ we cannot do better
than this order-of-magnitude estimate.

\subsection{Ekpyrotic conversion}
\label{subsectionEkpyroticConversion}

In this section, we analyze the case where the conversion of
entropy into curvature modes occurs during the ekpyrotic phase.
Indeed, if the background trajectory starts out slightly off
the ridge in the ekpyrotic potential, then after some time, the
field will naturally roll off down one side
\cite{Koyama:2007mg,Koyama:2007ag,Creminelli:2007aq}. For this
scenario, the non-gaussianity up to second order has been
analyzed by Koyama {\it et al.} \cite{Koyama:2007if} using the
$\d N$ formalism, with the result that $f_{NL}=-5/12 \, c_1^2,$
for the case where $\phi_1$ is the field that is frozen in at
late times, and where $c_1$ is the constant appearing in the
ekpyrotic potential (\ref{potential2field}). For
ekpyrotic conversion, the calculation is most easily performed,
and the result most easily expressed, in terms of the potential
(\ref{potential2field}), which is why we are adopting this
restricted form here. In working with a parameterized potential
like (\ref{potentialParameterized}), the bending of the
trajectory can be more complicated, in the sense that there can
be multiple turns, and one has to decide when to stop the
evolution. In this case, the results are strongly cut-off
dependent, and without a precisely defined model specifying the
subsequent evolution, it is impossible to make any generic
predictions. For kinetic conversion, this problem does not
arise, since the reflection potential is entirely unrelated to
the potential during the ekpyrotic phase. The bottom line is
that we will stick with the form (\ref{potential2field}) in
this section.

The $\d N$ formalism is particularly well suited to the case of
ekpyrotic conversion, as the background evolution is simple. In
fact, it turns out that by making the approximation that the
bending is instantaneous, it is very easy to find an approximate formula for
the non-linearity parameters at any chosen order in
perturbation theory. Below, we will summarize the calculation
of \cite{Koyama:2007if} and extend it to third order.
Subsequently, we will compare the approximate formula thus
obtained with the result from solving and integrating the
equations of motion numerically. In a sense, this provides a
check of the instantaneous bending approximation. At second
order in perturbation theory, it has been checked explicitly
that if the instantaneous bending approximation is relaxed,
then both methods (namely a numerical calculation using the $\d
N$ formalism, and a numerical calculation using the equations
of motion directly) agree to high precision in the value of
$f_{NL}$ that they predict, and moreover that the results are
in good agreement with the formula quoted above
\cite{Lehners:2008my}.

In order to implement the $\d N$ formalism, we have to
calculate the integrated expansion $N = \int H dt$ along the
background trajectory. Initially, the trajectory is close to
the scaling solution (\ref{ScalingSolution}). Then, we assume
that at a fixed field value $\d s_B$ away from the ridge, the
trajectory instantly changes course and rolls off along the
$\phi_2$ direction. At this point, the trajectory follows the
single-field evolution \be a(t)=(-t)^{2/c_2^2} \qquad \phi_2 =
\frac{2}{c_2} \ln(-t) + {\rm \, constant} \qquad \phi_1 = {\rm
constant}. \label{solutionSingleField}\ee It is now easy to
evaluate the integrated expansion, with the result that \be N =
-\frac{2}{c_1^2} \ln|H_B| + {\rm \, constant}, \ee where $H_B$
denotes the Hubble parameter at the instant that the bending
occurs. Note that all $c_2$-dependence has canceled out of the
formula above. At the end of the conversion process, we are
interested in evaluating the curvature perturbation on a
surface of constant energy density. But, in comoving gauge, the
curvature perturbation is equal to a perturbation in the
integrated expansion. Then, if we assume that the integrated
expansion depends on a single variable $\a$, we can write \be
\zeta= \d N = N_{,\a} \d \a + \frac{1}{2} N_{,\a\a} (\d \a)^2 +
\frac{1}{6} N_{,\a\a\a} (\d \a)^3. \ee In our example, we
indeed expect a change in $N$ to depend solely on a change in
the initial value of the entropy perturbation $\d s.$ Now, from
Eq. (\ref{entropyInitialCondition}), we know that $\d s_L
\propto 1/t \propto H,$ and hence we can parameterize different
initial values of the entropy perturbation by writing \be \d
s_L = \a H. \ee Note that since $\d s_L$ is gaussian, so is
$\a.$ With this identification, we have \be \d s= \a H +
\frac{\k_3 \sqrt{\e}}{8}(\a H)^2 +
\e(\frac{\k_4}{60}+\frac{\k_3^2}{80}-\frac{2}{5})(\a H)^3, \ee
so that at the fixed value $\d s = \d s_B,$ we have \be \a
\propto \frac{1}{H_B}. \ee Now we can immediately evaluate \be
N_{,\a} = N_{,H_B}\frac{d H_B}{d\a} = \frac{2}{c_1^2 \a},\ee
and, similarly \be N_{,\a\a} = -\frac{2}{c_1^2 \a^2} \qquad
N_{,\a\a\a} = \frac{4}{c_1^2 \a^3}. \ee In this way, with very
little work, we can estimate the non-linearity parameters \bea
f_{NL} &=&
\frac{5N_{,\a\a}}{6N_{,\a}^2} = -\frac{5}{12} c_1^2 \\
\tau_{NL} &=& \frac{36}{25}f_{NL}^2 = \frac{1}{4} c_1^4 \\
g_{NL} &=& \frac{25 N_{,\a\a\a}}{54 N_{,\a}^3} =
\frac{25}{108}c_1^4. \eea We are now in a position to compare
these estimates to the numerical results obtained by solving
and integrating the equations of motion (\ref{s3'}) and
(\ref{zeta3'}). In order to do this, we choose initial
conditions that are given by the scaling solution
(\ref{ScalingSolution}), except that we increase the initial
field velocity $|\phi_2'|$ by $0.1$ percent. This causes the
trajectory to eventually roll off in the $\phi_2$ direction,
and to quickly approach the single-field solution
(\ref{solutionSingleField}). The results for several values of
$c_1$ and $c_2$ are shown table \ref{Table1}, alongside the values
estimated by the $\d N$ formulae.

\begin{table}
\begin{tabular}{|c|c||c|c|c||c|c|c|}
  \hline
  $c_1$ & $c_2$ & $f_{NL,\delta N}$ & $\tau_{NL,\d N}$ & $g_{NL, \d N}$ & $f_{NL}$ & $\tau_{NL}$ & $g_{NL}$ \\ \hline
  10 & 10 & -41.67  & 2500 & 2315 & -39.95  & 2298 & 2591 \\
  10 & 15 & -41.67  & 2500 & 2315 & -40.45  & 2356 & 2813 \\
  10 & 20 & -41.67  & 2500 & 2315 & -40.62  & 2377 & 3030 \\
  15 & 10 & -93.75  & 12660 & 11720 & -91.01  & 11930 & 13100 \\
  15 & 15 & -93.75  & 12660 & 11720 & -92.11  & 12220 & 13830 \\
  15 & 20 & -93.75  & 12660 & 11720 & -92.49  & 12320 & 14440 \\
  20 & 10 & -166.7  & 40000 & 37040 & -162.5  & 38020 & 41320 \\
  20 & 15 & -166.7  & 40000 & 37040 & -164.4  & 38930 & 43170 \\
  20 & 20 & -166.7  & 40000 & 37040 & -165.1  & 39240 & 44490 \\
  \hline
\end{tabular}
\caption{\small Ekpyrotic conversion: the values of the non-linearity parameters estimated by the $\d N$ formalism compared to the numerical results obtained by directly integrating the equations of motion.}
\label{Table1}
\end{table}

It is immediately apparent that the general trend is accurately
captured by the $\d N$ formulae. However, one may notice that
the agreement is slightly less good at third order than at
second, and also, that the $\d N$ formulae tend to slightly
over-estimate $\tau_{NL}$ and slightly under-estimate $g_{NL}.$
But given the quickness of the $\d N$ calculation and the
complexity of the third order equations of motion, the
agreement is pretty impressive. Of course, the $\d N$ formula
was derived subject to the instantaneous bending approximation.
Without this approximation, we would expect a numerical scheme
that uses the $\d N$ formalism to yield results in close
agreement with our numerical results.

The general conclusion is that, contrary to $f_{NL},$ the sign
of $g_{NL}$ turns out to be always positive. Moreover, both
$\tau_{NL}$ and $g_{NL}$ scale very fast with increasing
equation of state $w_{ek} \sim c_1^2,$ and hence we can expect
future observations to be highly constraining for this type of
conversion (current observations are in fact already rather
constraining regarding $f_{NL}$
\cite{Komatsu:2008hk,Smith:2009jr}).

Finally, we should note that a conversion mechanism similar to
ekpyrotic conversion (in the sense that the conversion is
assumed to happen during the ekpyrotic phase) has been
considered in the new ekpyrotic models of Buchbinder {\it et
al.}
\cite{Buchbinder:2007ad,Buchbinder:2007tw,Buchbinder:2007at},
except that the roll-off from the ridge is expected to be
caused by a feature in the potential rather than by initial
conditions that are slightly off-centered. We would simply like
to note that for these models, there is in fact considerably
more flexibility, as the ekpyrotic phase is followed by a ghost
condensate phase. The details of how the transition to the
ghost condensate phase occurs will determine what one expects
for the non-linearity parameters $f_{NL},\tau_{NL},g_{NL}.$ For
example, it is conceivable that the ekpyrotic phase might be
followed by a kinetic phase during which the ghost condensate
starts dominating. In that case, one would expect the results
to be more closely aligned with those presented in the previous
section. In the absence of a concrete model, we must postpone
making definite predictions for these models.

\section{Discussion}

The analysis of the distribution of density fluctuations,
either via the CMB or via large-scale structure surveys,
currently offers the best prospects for gaining information
about the time around the big bang. In ekpyrotic and cyclic
models, the pattern of these fluctuations is imprinted during a
slowly contracting ekpyrotic phase preceding the current
expanding phase. We have restricted our analysis to the most
robust mechanism to date by which the density fluctuations can
arise in these models, namely the entropic mechanism in which entropy
perturbations are generated first, and are then converted into
curvature perturbations just before the big bang. The
conversion can happen in at least two distinct ways, either
directly during the ekpyrotic phase or during the kinetic
energy dominated phase that leads up to the big crunch - big
bang transition. In this paper, we have assumed that the
dominant amount of conversion occurs before the big crunch, and
that the resulting density perturbations evolve essentially
unchanged up to the time of nucleosynthesis, so as to become
the `primordial' density perturbations.

For both conversion modes, we have found that the combined consideration of the non-linearity
parameters $f_{NL}$ (and by extension $\tau_{NL}$) and $g_{NL}$
results in a distinct observational imprint, that should
enable one to select or rule out these models on observational
grounds in the foreseeable future. In short, for ekpyrotic
conversion, $f_{NL}$ is always negative and typically of ${\cal
O}(10-100)$, while $g_{NL}$ is always positive and typically of ${\cal
O}(10^3-10^4)$. For kinetic conversion, on the other hand,
$f_{NL}$ can have either sign and is typically of ${\cal
O}(10),$ while $g_{NL}$ is typically negative and of ${\cal
O}(1000).$ Thus, typical values are substantially different from the predictions of simple single-field inflationary models, and the local form of the produced non-gaussianity should also easily distinguish ekpyrotic models from single-field inflationary models with non-canonical kinetic terms. The comparison with multi-field inflationary models is more subtle, as some of those models can allow for virtually any values of $f_{NL}$ and $g_{NL}$. However, should the observed values happen to lie in the ranges predicted in this paper, then ekpyrotic models will provide strong candidates for a model of the early universe due to the correlation between the values of $f_{NL}$ and $g_{NL}.$ It will certainly  be exciting to compare the predictions derived here to
the results of future observations.

\vspace{2cm}

\noindent {\it Acknowledgements} We would like to thank E. Komatsu, K. Koyama, D. Langlois, P. Steinhardt for stimulating and valuable discussions.

\vspace{1cm}

\section*{Appendix - Useful formulae}

\bea
u^0 &=& \frac{dt}{d\t} = 1-A^{(1)}-A^{(2)}+\frac{3}{2}A^{(1)2}-A^{(3)}+3A^{(2)}A^{(1)}-\frac{5}{2}A^{(1)3} \\ u^i &=& 0,
\eea
where $A$ is the lapse function defined by $g_{00} \equiv -(1+2A).$

\bea \dot{\z}_i^{(3)} &=&
\z_i^{(3)'}-A\z_i^{(2)'}-A^{(2)}\z_i^{(1)'}+\frac{3}{2}A^2\z_i^{(1)'}
\\ \ddot{\z}_i^{(3)} &=&
\z_i^{(3)''}-2A\z_i^{(2)''}-A'\z_i^{(2)'}-2A^{(2)}\z_i^{(1)''}-A^{(2)'}\z_i^{(1)'}+4A^2\z_i^{(1)''}+4AA'\z_i^{(1)'}
\eea

\bea
\delta e_{\sigma}^I \tackl = \tackr \frac{1}{\sib'} \left(\delta s'+\bar \theta' \delta \sigma \right) \bes^I
\label{deltaesigma} \\
\delta e_{s}^I \tackl = \tackr -\frac{1}{\sib'} \left(\delta s'+\bar \theta' \delta \sigma \right) \besi^I
\label{deltaes}
\eea

\bea
\delta e_{\s}^{I(2)}&=&-\frac{\besi^I}{2 \sb^{'2}} \left( \delta s'+ \tb' \delta \s \right)^2+\frac{\bes^I}{\sb'}\left[-\frac{1}{\sb'}\left(\delta \s'-\tb' \delta s \right)\left( \delta s'+ \tb' \delta \s  \right)+\tb' \left( \delta \s^{(2)}-\frac{1}{2 \sb'}\delta s \delta s' \right) \right.
\cr
&+&\left.
\delta s^{(2)'}+\left(\frac{\delta \s}{\sb'}\left(\delta s'+ \frac{\tb'}{2} \delta \s  \right) \right)' \right]
\label{delta2esigma}
\eea

\bea
\delta e_{s}^{I(2)}&=&-\frac{\bes^I}{2 \sb^{'2}} \left( \delta s'+ \tb' \delta \s \right)^2-\frac{\besi^I}{\sb'}\left[-\frac{1}{\sb'}\left(\delta \s'-\tb' \delta s \right)\left( \delta s'+ \tb' \delta \s  \right)+\tb' \left( \delta \s^{(2)}-\frac{1}{2 \sb'}\delta s \delta s' \right) \right. \label{esi2}
\cr
&+&\left.
\delta s^{(2)'}+\left(\frac{\delta \s}{\sb'}\left(\delta s'+ \frac{\tb'}{2} \delta \s  \right) \right)' \right]
\label{delta_es_2}
\eea

\bea
\delta \dot \s^{(1)}=u^{0 (1)} \sb' +\delta \s'-\tb' \delta s
\label{dsigma1}
\eea

\bea
\delta \dot \s^{(2)}&=&u^{0 (2)} \sb' +u^{0(1)} \delta \s' + \d \s^{(2)'}  \nn \\
&-&u^{0 (1)} \tb' \d s-\tb' \d^{(2)} s +(\bar V_{,ss}+3 \tb^{'2})\frac{\d s^2}{2 \sb'}-\frac{\bar V_{,\s}}{2 \sb^{'2}} \d s\d s' +\frac{\tb'}{\sb^{'2}} \d s\d \epsilon
\label{dsigma2}
\eea

\bea
\left(\frac{\dot \r}{\dot \s} \right)^{(1)}=\left( \frac{\d \r'}{\sb'}-\frac{\rb'}{\sb^{'2}}\d \s'\right)+\frac{\rb'}{\sb^{'2}}\tb' \d s
\label{rsi1}
\eea

\bea \left(\frac{\dot \r}{\dot \s} \right)^{(2)} &=& \frac{\d
\r'}{\sb'} \left[ \left( \frac{\d \s'}{\sb'} \right)^2-\frac{\d
\s'}{\sb'} \frac{\d \r'}{\rb'}
\right]+\frac{\r^{(2)'}}{\sb'}-\frac{\rb'}{\sb^{'2}}\d
\s^{(2)'} \cr &+&  \frac{\d \r'}{\sb^{'2}} \left[\tb' \d
s^{(2)}-(\bar V_{,ss}+\tb'^2)\frac{\d s^2}{2 \sb'}+ \frac{\bar
V_{,\s}}{2 \sb^{'2}}\d s \d s'-\frac{\tb'}{\sb^{'2}}\d s \d \e
-\frac{2 \tb' \d s \d \s'}{\sb'}+\frac{\d \r'}{\rb'}\tb' \d s
\right] \label{rsi2} \eea

In comoving gauge and on large scales
\bea \d V_{,s} &\approx& \bar V_{,ss}\d s -\frac{\bar V_{,\s}}{\sb'}\d s' \\ \d V_{,s}^{(2)} &\approx& \frac{1}{2}\bar V_{,sss}\d s^2 +
\bar V_{,ss}\d s^{(2)} - \frac{3}{2\sb'}\bar V_{,s\s}\d s \d s' -\frac{\bar V_{,s}}{2\sb^{'2}}\d s'^2 - \frac{\bar V_{,\s}}{\sb'}(\d s^{(2)'}+\frac{\tb'}{2\sb'}\d s \d s')\\ \d V_{,ss} &\approx& \bar V_{,sss}\d
s -2\frac{\bar V_{,s\s}}{\sb'}\d s' \\ \d V_{,ss}^{(2)} &\approx& \frac{1}{2}\bar V_{,ssss}\d s^2 + \bar V_{,sss}\d s^{(2)} -\frac{5}{2\sb'}\bar V_{,ss\s} \d s \d s' +\frac{\bar V_{,\s\s}-\bar V_{,ss}}{\sb^{'2}}\d s'^2 \nn \\ && - 2\frac{\bar V_{,s\s}}{\sb'}(\d s^{(2)'}+\frac{\tb'}{2\sb'}\d s \d s') \\ \d
\dot{\th}^{(2)} &\approx& \frac{\bar V_{,\s}}{\sb'^2}\d s^{(2)'} -
\frac{1}{\s'}(\bar V_{,ss}+\tb'^2)\d s^{(2)} -\frac{\tb'}{2\sb^{'3}}\d s'^2 +
\frac{1}{2\sb^{'2}}\left(4\frac{\tb'\bar V_{,\s}}{\sb'}-3\tb''+9H\tb'\right)\d s' \d s \nn \\ && +
\frac{1}{2\sb^{'2}}\left(-\sb'\bar V_{,sss}+3\bar V_{,ss}\tb'+3\tb^{'3}\right)\d s^2 \\
A &\approx& -2\frac{\tb'}{\sb'} \d s \\ A^{(2)} &\approx& -2\frac{\tb'}{\sb'}\d s^{(2)} +\frac{1}{\sb'^2}(\bar V_{,ss} + 6 \tb'^2)\d s^2 -\frac{\bar V_{,\s}}{\sb'^3}\d s \d s' \eea
Finally, the constraint equations derived in \cite{Langlois:2006vv} imply, in the comoving gauge and on large scales \bea \Theta^{(1)} &\approx& 0\,, \\ \Theta^{(2)} &\approx& 0. \eea

\vspace{2cm}

\bibliography{Trispectrum}

\end{document}